\documentclass[10pt, journal]{IEEEtran}

\usepackage{cite}
\usepackage{amsmath,amssymb,amsfonts}
\usepackage{graphicx}
\usepackage{textcomp}
\usepackage{xcolor}
\usepackage{caption}
\usepackage{algorithm}
\usepackage[noend]{algpseudocode}
\usepackage{scalerel}
\usepackage{subcaption}
\usepackage{cuted}
\usepackage[nolist,nohyperlinks]{acronym}

\usepackage{tikz, circuitikz}
\usetikzlibrary{arrows,decorations.pathmorphing,backgrounds,positioning,fit,petri,quotes}
\usepackage{import}

\usetikzlibrary{svg.path}
\definecolor{orcidlogocol}{HTML}{A6CE39}
\tikzset{
  orcidlogo/.pic={
    \fill[orcidlogocol] svg{M256,128c0,70.7-57.3,128-128,128C57.3,256,0,198.7,0,128C0,57.3,57.3,0,128,0C198.7,0,256,57.3,256,128z};
    \fill[white] svg{M86.3,186.2H70.9V79.1h15.4v48.4V186.2z}
                 svg{M108.9,79.1h41.6c39.6,0,57,28.3,57,53.6c0,27.5-21.5,53.6-56.8,53.6h-41.8V79.1z M124.3,172.4h24.5c34.9,0,42.9-26.5,42.9-39.7c0-21.5-13.7-39.7-43.7-39.7h-23.7V172.4z}
                 svg{M88.7,56.8c0,5.5-4.5,10.1-10.1,10.1c-5.6,0-10.1-4.6-10.1-10.1c0-5.6,4.5-10.1,10.1-10.1C84.2,46.7,88.7,51.3,88.7,56.8z};
  }
}
\newcommand\orcidicon[1]{\href{https://orcid.org/#1}{\mbox{\scalerel*{
\begin{tikzpicture}[yscale=-1,transform shape]
\pic{orcidlogo};
\end{tikzpicture}
}{|}}}}

\usepackage{hyperref}
\hypersetup{
colorlinks=true, %
linktoc=all,     %
linkcolor=black,  %
}

\def\BibTeX{{\rm B\kern-.05em{\sc i\kern-.025em b}\kern-.08em
    T\kern-.1667em\lower.7ex\hbox{E}\kern-.125emX}}

\DeclareMathOperator*{\argmin}{arg\,min}

\title{Energy-Constrained Information Storage on Memristive Devices in the Presence of \\ Resistive Drift}

\author{\IEEEauthorblockN{Waleed El-Geresy \orcidicon{0000-0002-4016-6078},~\IEEEmembership{Graduate Student Member, IEEE},
 Christos Papavassiliou \orcidicon{0000-0002-8003-2146},~\IEEEmembership{Senior Member, IEEE}, and Deniz Gündüz \orcidicon{0000-0002-7725-395X},~\IEEEmembership{Fellow, IEEE}}
 \thanks{Authors are affiliated with the Department of Electrical and Electronic Engineering, Imperial College London. Emails: [waleed.el-geresy15, c.papavas, d.gunduz] @imperial.ac.uk. This work was supported by the EPSRC under the DTP 2016-2017 (EP/N509486/1), DTP 2018-2019 (EP/R513052/1), SONATA (EP/W035960/1) and FORTE (EP/R024642/1) projects.}}

\begin{document}

\maketitle

\thispagestyle{plain}
\pagestyle{plain}

\begin{abstract}

    In this paper, we examine the problem of information storage on memristors affected by resistive drift noise under energy constraints. We introduce a novel, fundamental trade-off between the information lifetime of memristive states and the energy that must be expended to bring the device into a particular state. We then treat the storage problem as one of communication over a noisy, energy-constrained channel, and propose a \ac{JSCC} approach to storing images in an analogue fashion. To design an encoding scheme for natural images and to model the memristive channel, we make use of data-driven techniques from the field of deep learning for communications, namely \ac{DeepJSCC},  employing a generative model of resistive drift as a computationally tractable differentiable channel model for end-to-end optimisation. We introduce a modified version of \ac{GDN}, a biologically inspired form of normalisation, that we call \ac{cGDN}, allowing for conditioning on continuous channel characteristics, including the initial resistive state and the delay between storage and reading. Our results show that the delay-conditioned network is able to learn an energy-aware coding scheme that achieves a higher and more balanced reconstruction quality across a range of storage delays.

\end{abstract}


\begin{acronym}[]%
        \acro{AWGN}{additive white Gaussian noise}

        \acro{BN}{batch normalisation}

        \acro{cBN}{conditional batch normalisation}
        \acro{cGDN}{conditional GDN}
        \acro{cGAN}{conditional generative adversarial network}
        \acro{CPU}{central processing unit}

        \acro{DeepJSCC}{deep joint source-channel coding}

        \acro{ECM}{electrochemical metallization}
        
        \acro{GAN}{generative adversarial network}
        \acro{GDN}{generalised divisive normalisation}
        \acro{GMSM}{Generalised Metastable Switch Model}
        \acro{GPU}{graphics processing unit}

        \acro{icGDN}{inverse conditional GDN}
        \acro{iGDN}{inverse GDN}
        
        \acro{JSCC}{joint source-channel coding}

        \acro{MIM}{Metal-Insulator-Metal}
        \acro{MSE}{mean squared error}

        \acro{PCM}{phase change memory}
        \acro{PSNR}{peak signal-to-noise ratio}

        \acro{RAM}{random access memory}
        \acro{RRAM}{resistive random access memory}

        \acro{SNR}{signal-to-noise ratio}
        \acro{SGD}{stochastic gradient descent}

        \acro{VQA}{Visual Question Answering}
\end{acronym}

\acresetall

\section{Introduction}
\label{sec:introduction}
\begin{figure}[t]
    \centering
    \includegraphics[width=\linewidth]{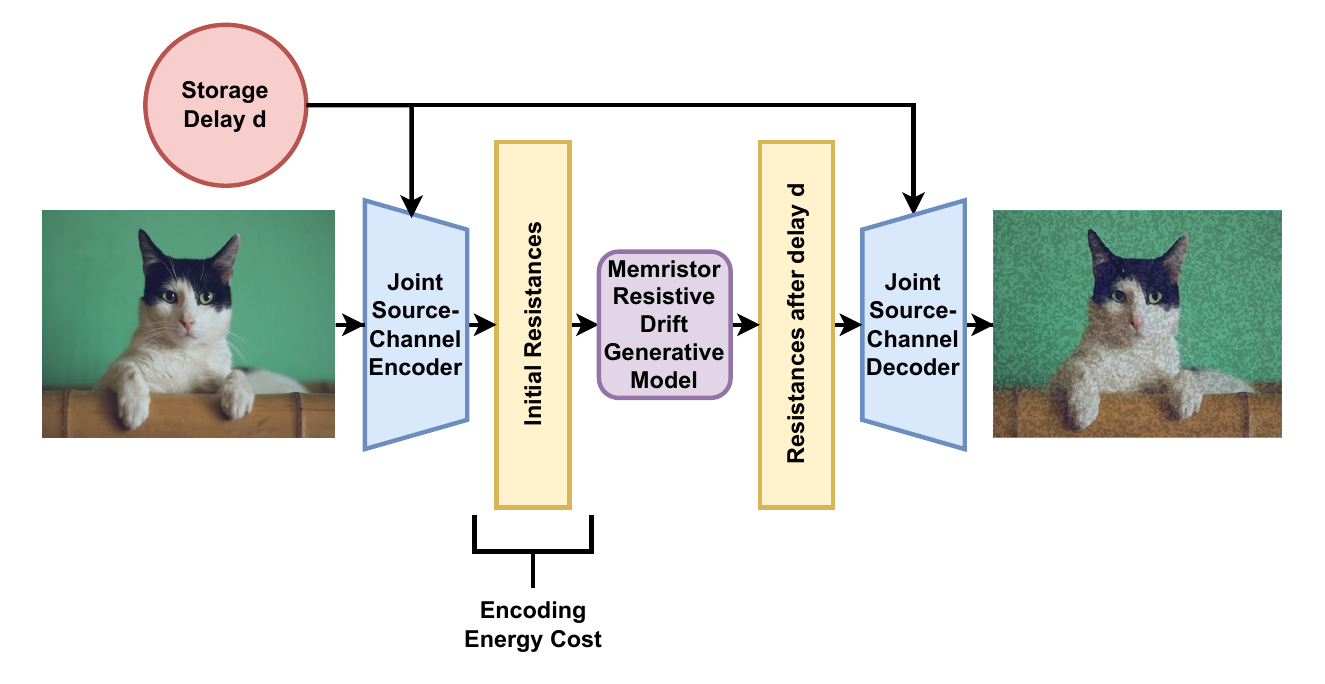}
    \caption{In traditional wireless communications information is transmitted through space, whereas, in the case of storage, transmission is through time. We present a \ac{DeepJSCC} approach to learning to store semantic information on memristive devices, under energy constraints, in the presence of resistive drift noise.
    }
    \label{fig:information_storage_diagram}
\end{figure}

\IEEEPARstart{I}{nformation} storage is fundamentally a problem of communication. In contrast to the more familiar wireless communication problem, where the challenge is to transmit information from one \textit{spatial} location to another, information storage is the problem of transmitting information from one \textit{temporal} point to another (see Figure~\ref{fig:information_storage_diagram}).
The treatment of a storage medium as a noisy communication channel, and the use of channel coding in the form of error correction codes (sometimes also incorporating source coding \cite{leeCombinedSourceChannel1999}) to reliably recover stored information has been studied extensively for traditional storage media.
For traditional storage media, there is a trade-off between the longevity/robustness of the stored information and the number of bits that can be stored. For a storage medium that exhibits degradation over time (notably \ac{RAM}), there exists a trade-off between the duration we wish our information to be reliably recoverable for and the storage density (the number of levels we use to store information) \cite{larriveeSolidStateDrive2015}, with a longer period between storage and recovery leading to a degradation in the number of distinct, recoverable levels and hence the amount of information that can be stored.

To go beyond the limits of traditional storage media in terms of storage density or operational power requirements, novel storage technologies have been proposed. One class of devices that has been proposed for this purpose is the \textit{memristor}.
Memristors are electronic devices with a semi-volatile resistive state (memory) that can be modified through a driving voltage or current. The memristor was originally hypothesised by Chua \cite{chuaMemristorMissingCircuit1971} as the missing fundamental passive circuit element in a ``periodic table'' of passive electronic devices. Later, the definition and scope of the term ``memristor'' expanded to encompass a variety of resistive switching technologies \cite{chuaMemristiveDevicesSystems1976, chuaIfItsPinched2014, chuaResistanceSwitchingMemories2011}. %
Notable types of devices that have been described as ``memristive'' include \ac{PCM} devices \cite{engelCapacityOptimizationEmerging2015} and thin film metal oxide devices \cite{strukovMissingMemristorFound2008}. Indeed, \ac{PCM} has already seen widespread application in optical storage, though its resistive storage properties are still an emerging technology. Possible uses of memristors go beyond information storage into the realm of \textit{neuromorphic computing} \cite{sungPerspectiveReviewMemristive2018} - computing inspired by principles of human brain. One driving factor behind this use case is their ability to co-locate the processing and storage of information, thus potentially enabling computation to occur in a distributed and massively parallelised way \cite{meadNeuromorphicElectronicSystems1990}, with processing happening in-place, circumventing the so-called Von Neumann bottleneck \cite{vonneumannFirstDraftReport1993}.

It is essential to characterise optimal methods for information storage on memristors to exploit as much of their available capacity as possible, taking into account non-idealities inherent in the devices. Information storage can be addressed from a variety of perspectives: memristive devices can be treated as binary storage devices \cite{rumseyCapacityConsiderationsData2019}, multilevel storage devices \cite{wangMultiStateMemristorsTheir2022}, or even continuous analogue storage devices \cite{zarconeAnalogCodingEmerging2020}.

The first step towards designing optimal storage mechanisms is to understand and model the characteristics of the noisy storage channel. Different types of noise affect memristive devices, such as programming noise, sneak-path current noise, read noise, device parameter heterogeneity due to manufacturing differences, and finally noise associated with the metastability of the resistive state, including resistive drift and burst noise (random telegraph noise). Most of these are time-invariant, spatial noise processes, whereas resistive drift is distinct in that it is dependent on the delay between writing a value to the device and subsequently attempting to recover it accurately.
The problem of resistive drift has been noted in a number of works - including the suggestion that the modelling of memristors should include an autonomous state evolution component \cite{carbajalMemristorModelsMachine2015}. In thin film metal oxide memristive devices the resistive state decays stochastically over time due to the phenomenon of resistive drift, until an equilibrium resistance is attained and all the stored information is eventually lost \cite{ielminiModelingUniversalSet2011}. The longer the delay between storage and recovery, the less likely it will be to recover the original information reliably, while the use of a wider range of initial resistances makes it more likely that some of the initial stored information can be preserved. Hence, the limits of information storage can be analysed in terms of the information capacity of a device conditioned upon the delay.

Sneak path current noise and the resultant capacity of an array of (binary) memristive devices when used in a crossbar circuit was studied theoretically using a graph theoretic approach in \cite{rumseyCapacityConsiderationsData2019}. A practical neural coding scheme for decoding information stored in the presence of sneak path currents in memristive crossbar arrays is proposed in \cite{sunBeliefPropagationBased2021}, using belief propagation decoding and polar codes, designed by a genetic algorithm. The capacity of \ac{PCM} memristive devices used in the multi-level storage domain, in the presence of programming noise, has been estimated using discretisation and the Blahut-Arimoto algorithm \cite{engelCapacityOptimizationEmerging2015, engelOpportunitiesAnalogCoding2017}. Following on from this initial capacity consideration, deep learning techniques for \ac{JSCC} - namely a \ac{DeepJSCC} autoencoder - were used to address the problem of information storage in \ac{PCM} memristors in the presence of programming noise \cite{zarconeJointSourcechannelCoding2018, zhengErrorResilientAnalogImage2018, zarconeAnalogCodingEmerging2020} in both a theoretical, and subsequently realistic, setting.

\ac{JSCC} aims to optimise the quality of the recovered information by jointly considering the problems of (lossy) compression and mitigation of the channel noise effects, in an end-to-end fashion \cite{gunduzJointSourceChannel2024}. 
Separate source and channel coding techniques dominate almost all modern communication systems, despite their sub-optimality in short block-length regimes. This is due to both the modularity they provide, as well as to the lack of superior, practical \ac{JSCC} schemes. As a result, limited attention has been given to designing joint schemes. This trend has recently changed with the advent of deep learning aided JSCC schemes pioneered in \cite{bourtsoulatzeDeepJointSourceChannel2019}.
\ac{DeepJSCC} refers to the use of deep neural networks to parameterise the joint encoder and decoder to perform \ac{JSCC}, trained for particular kinds of sources, such as images or videos, over specific communication channels \cite{bourtsoulatzeDeepJointSourceChannel2019, tungDeepWiVeDeepLearningAidedWireless2022, tungDeepJSCCQConstellationConstrained2022, xuDeepJointSourceChannel2023}.

In this work, we focus on resistive drift in the context of energy-constrained information storage, making use of \ac{DeepJSCC} to optimise the end-to-end distortion of natural images encoded and stored on memristors. In this context, the communication channel is simply a conditional probability distribution, which depends on the delay between storage and recovery.
The delay can be seen as a channel condition analogous to the \ac{SNR} in wireless \ac{AWGN} channels.
Another important factor affecting the reliability of the storage channel is the available energy budget for storing information. While a number of works have studied the energy requirements of bringing memristors into particular resistive states \cite{fuBioinspiredBiovoltageMemristors2020, ielminiSizeDependentRetentionTime2010, ielminiModelingUniversalSet2011, weiDemonstrationHighdensityReRAM2011}, the energy-distortion trade-off has thus far, to the best of our knowledge, not been explicitly considered in this context.
We will explore this fundamental trade-off between the energy budget for storage, and the reliability of the memristor as a storage channel (its capabilities for retention of information).
We will then consider the impact of the energy consumption on the storage problem, presenting it as one of optimising a trade-off between the energy consumption and the quality of the recovered information in terms of the prescribed end-to-end distortion measure under different storage delays. In particular, we will explore novel techniques for enabling a \ac{DeepJSCC} network to adapt to the resistive drift channel conditions (initial resistance and delay) for a given energy budget.

\subsection{Contributions}

The contributions of this paper can be summarised as follows:
\begin{enumerate}
    \item For the first time in the literature, we expose and investigate the trade-off between the energy consumption, information storage lifetime, and recovery distortion in memristive devices, formulating this as a \ac{JSCC} problem.
    
    \item We present a deep learning based solution for the presented \ac{JSCC} problem. In particular, we introduce a novel technique based on conditioning a deep joint source channel encoder/decoder pair on the storage delay - through introducing a modified form of the \ac{GDN} transformation that we call \ac{cGDN}. To the best of our knowledge, this is the first time that delay-dependent encoding and decoding has been explored in the context of data storage.
\end{enumerate}

The rest of this paper is organised as follows. The treatment of resistive drift in the context of information storage is formalised in Section~\ref{sec:problem_formulation}.
In Section~\ref{sec:energy_resistance_level_tradeoff}, we introduce and characterise a cost function that yields the energy requirements for programming different states in memristive devices. Following this, in Section~\ref{sec:autoencoder_storage}, we introduce a deep learning based approach to the energy-constrained storage problem, conditioned on the delay to information recovery.

\textbf{Notation} We use lowercase \(r\) to denote resistance values, and \(d\) to denote delay values. We use the notation \(\bar{y}\) to denote a version of a given variable \(y\) to which a corresponding data normalisation has been applied. \(N_x(\cdot)\) is used to denote a particular data normalisation transform. Note that the form of the normalisation will differ between variables.

\section{Problem Formulation}
\label{sec:problem_formulation}

We consider storage of natural images on memristive devices that are subject to resistive drift, which depends on the delay between the time of storage and recovery, \(T\).
We denote the image to be stored by the random variable \(X \in \mathbb{R}^{H\times W\times C}\), where \(H\), \(W\), and \(C\) denote the height, width and number of colour channels of the image, respectively. We store a representation of \(X\), denoted by the random variable \(M \in \mathbb{R}^n\), on an array of \(n\) memristor devices for the duration \(T \sim p_T(t)\) with support \([t_{\text{min}}, t_{\text{max}}]\). This can also account for the case of a fixed delay, where the distribution is a degenerate distribution, with probability mass only at that particular delay. Our goal is to recover an estimate of \(X\), denoted by \(\hat{X}\), from the noisy version of \(M\), denoted by \(\hat{M}\), which is the output of the storage channel (a conditional probability distribution) after delay \(T\), such that the distortion between \(X\) and \(\hat{X}\) is small according to a given measure.

In order to obtain the representation \(M\), we first encode the image using an encoder mapping \(E_{\theta_e} : \mathbb{R}^{H \times W \times C} \rightarrow \mathbb{R}^{n}\), which is parameterised by \(\theta_e\), and maps \(X\) to the memristive random vector \(M\). This is passed through conditional probability distribution representing the delay-conditioned noisy storage channel \(P(\hat{M}|T, M)\). We subsequently decode the channel output, \(\hat{M}\),  using the decoder mapping \(D_{\theta_d}: \mathbb{R}^{n} \rightarrow \mathbb{R}^{H \times W \times C}\) to produce a reconstruction \(\hat{X}\).
For the encoder and the decoder mappings, if the delay is known then their parameters will be conditional upon it.

When there is a cost associated with different codewords due to a limited energy budget for transmission (storage), the channel inputs \(M\) will be constrained.
In particular, we have

\begin{align}
    \mathbb{E}_{M}[\mathcal{E}(m)] \leq B
    \label{eq:energy_budget_condition}
\end{align}

where \(\mathcal{E}(\cdot)\) is the energy cost function, mapping a given input codeword (memristive value) \(M=m\) to an associated energy cost, and the expectation is over the distribution of possible stored codewords induced by the encoder mapping and the input image distribution.

The channel itself is a memristive storage channel, whose output is a conditional random variable, conditioned on the value of the delay, \(T\), and the given initial memristor resistance values, \(M\), and is denoted \(\hat{M}(T, M)\). Over the various delay conditions \(T \in [t_{\text{min}}, t_{\text{max}}]\), the information storage problem presents as a family of communication channels. The storage channel is similar in nature to a fading channel, where an increasing delay between the time of storage and recovery leads to diminishing channel quality (and consequently, achievable end-to-end distortion).

Our problem consists of minimising the expected end-to-end distortion between \(X\) and \(\hat{X}\), by choosing appropriate encoding and decoding functions.
In this work, these functions will be modelled as deep neural networks.

We will focus on the \ac{MSE} as the distortion measure to be optimised, although the mehtods and results can easily be generalised to any other differentiable loss function.
The distortion to optimise, \(l\), is equivalent to the Frobenius norm between the image and its reconstruction, i.e. \(l(X, \hat{X}) = \left|\left|X, \hat{X}\right|\right|_F\).
In addition, we will use the \ac{PSNR}, \(l_{\text{PSNR}}\), as a measure of the performance:

\begin{align}
    l_{\text{PSNR}}(X, \hat{X}) = 10 \cdot \log_{10}{\frac{1}{l(X, \hat{X})}}.
    \label{eq:psnr}
\end{align}

Thus, to summarise, our energy-constrained optimisation problem is that of finding the optimal parameters \(\theta_e^*\) and \(\theta_d^*\) that satisfy the following:

\begin{align}
    \{\theta_e^*, \theta_d^*\} = \argmin_{\theta_e , \theta_d} \mathbb{E}_{T, X}\left[ l(X, \hat{X}) \right] \\
    \text{where }
    \begin{cases}
    \mathcal{E}(E_{\theta_e}(X)) \leq B, \\
    \hat{X} = D_{\theta_d}(\hat{M}|(T, E_{\theta_e}(X)))
    \end{cases}
    \label{eq:objective_function}
\end{align}

We note that the average power constraint is computed over all codewords, rather than a single codeword, as is common in the traditional \ac{DeepJSCC} setting. This is to ensure that a trade-off between the energy expended in storing codewords on the memristors for different delays can be exploited, with more or less energy being allocated to enable storage, dependent on the quality of the storage channel for the given delay.

In Section~\ref{sec:energy_resistance_level_tradeoff}, we explore existing results from the literature to derive a parameterised energy cost function, \(\mathcal{E}\), that defines the energy consumption for a given channel input distribution in the context of the problem of resistance drift in memristors.

\section{Energy-Retention Trade-off}
\label{sec:energy_resistance_level_tradeoff}
A trade-off has been noted in the memristor literature between the total energy of a pulse needed for programming a device and the subsequent dynamic range or retention time. Filament size in tantalum oxide thin film memristors has been observed to impact the magnitude of the high and low resistive states \cite{weiDemonstrationHighdensityReRAM2011}. Fu et al. reported a linear relationship between the applied pulse width and the mean retention time for bio-voltage memristors (\ac{ECM} devices catalysed by protein nanowires) \cite{fuBioinspiredBiovoltageMemristors2020}. Ielmini examined the retention time as a function of the size of resistive filaments for \(NiO\) thin-film memristive devices, giving the final size of the resistive filaments as a function of the input signal pulse width applied to the devices \cite{ielminiSizeDependentRetentionTime2010}. Further to this, pulsing experiments performed on \(H_fO_x\) \ac{MIM} memristive devices were used to derive a generic inverse proportionality relationship between the compliance current and the final resistance of the memristive state achieved in \cite{ielminiModelingUniversalSet2011}.
If different memristive states are associated with different energy costs, then different storage patterns (input distributions) will be associated with different associated energy costs. If the dynamic range of the device
is decreased, this may result in a less reliable storage channel and a subsequent decrease in the information capacity (from a channel coding perspective) or an increase in the lower bound on the end-to-end distortion (from a \ac{JSCC} perspective).

Based on these findings, we have derived a generic relationship between the initial resistance (assumed to be the off or RESET state) of the devices and the energy needed to bring them to a particular state. This is taken into account in training the encoder and decoder functions through the introduction of a regularisation term in the loss function to represent the energy cost.

\subsection{Derivation of the Energy Cost Function}

When programming a memristive device, we can modulate either the current pulse width, or its magnitude, to modify the final switching state. As was explained above, it has been observed that for various kinds of memristive device:

\begin{enumerate}
    \item \textbf{For a fixed current magnitude}: there is a proportional relationship between the final conductance state of metal oxide memristive devices and the width (duration) of current pulses used to program the devices \cite{ielminiSizeDependentRetentionTime2010}.
    \item \textbf{For a fixed pulse width}: there is an inversely proportional relationship between the compliance current magnitude and the final resistance state \cite{ielminiModelingUniversalSet2011}.
\end{enumerate}

We use these two principles in the following discussion to derive a generic trade-off between the energy applied to the device, through the application of a current pulse of a given magnitude and duration, and the final resistive state achieved.
We will make use of the simplifying approximation that the amplitude of the current in an applied signal used to program the devices is constant and equal to its maximum (compliance) value throughout.
We begin by defining the relationship between a given compliance current magnitude and the final resistance attained.

Let \(R(\tau)\) (\(G(\tau)\)) be the resistance (conductance) at time \(\tau\) during the pulse. Let \(\tau_{\text{final}}\) be the (fixed) length of the pulse, required to bring the resistance to \(R(\tau_{\text{final}})\) (or the conductance to \(G(\tau_{\text{final}})\)).
Given that the total change in the conductance over the duration of the pulse is equal to \(G(\tau_\text{final}) - G(0)\), based on our assumption that the change in the conductance is linearly proportional to the duration of an applied current pulse, we can say that the proportion of this change effected is linearly proportional to the time \(\tau\) elapsed since the start of the pulse (a first order approximation).
Noting that \(R>0\), we can express \(R(\tau)\) as follows:
\begin{align}
    G(\tau) &= G(0) + \frac{\tau}{\tau_{\text{final}}} \cdot \big(G(\tau_{\text{final}})-G(0)\big) \notag, \\
    \notag \frac{1}{R(\tau)} &= \frac{1}{R(0)} + \frac{\tau}{\tau_{\text{final}}}\left(\frac{1}{R(\tau_{\text{final}})} - \frac{1}{R(0)}\right), \\
     R(\tau) &= \frac{1}{\left|\frac{1}{R(0)} + \frac{\tau}{\tau_{\text{final}}}\left(\frac{1}{R(\tau_{\text{final}})} - \frac{1}{R(0)} \right)\right|}
    \label{eq:resistance_as_func_time}
\end{align}

\begin{figure*}
    \centering
    \includegraphics[width=0.45\linewidth]{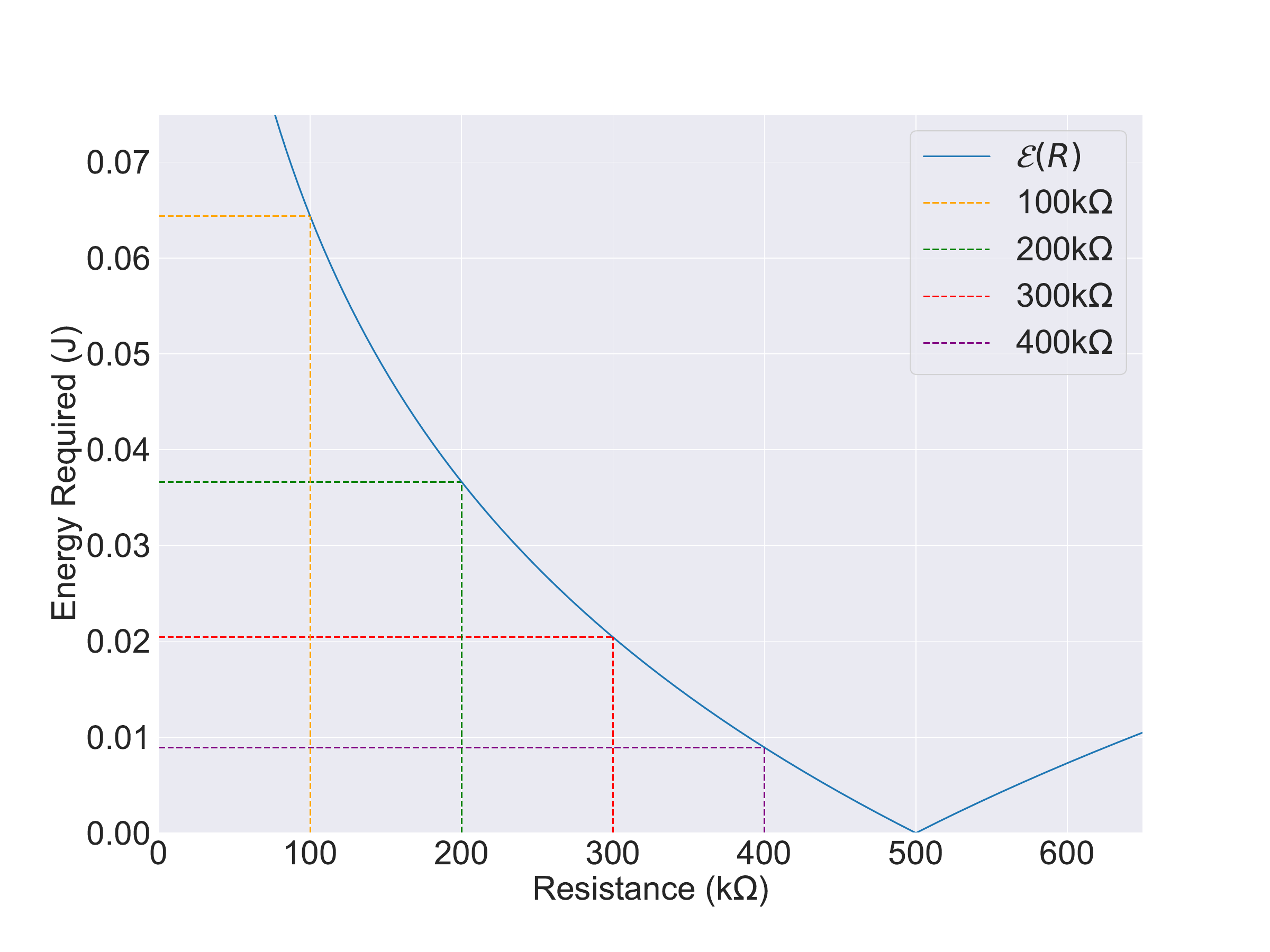}
    \includegraphics[width=0.45\linewidth]{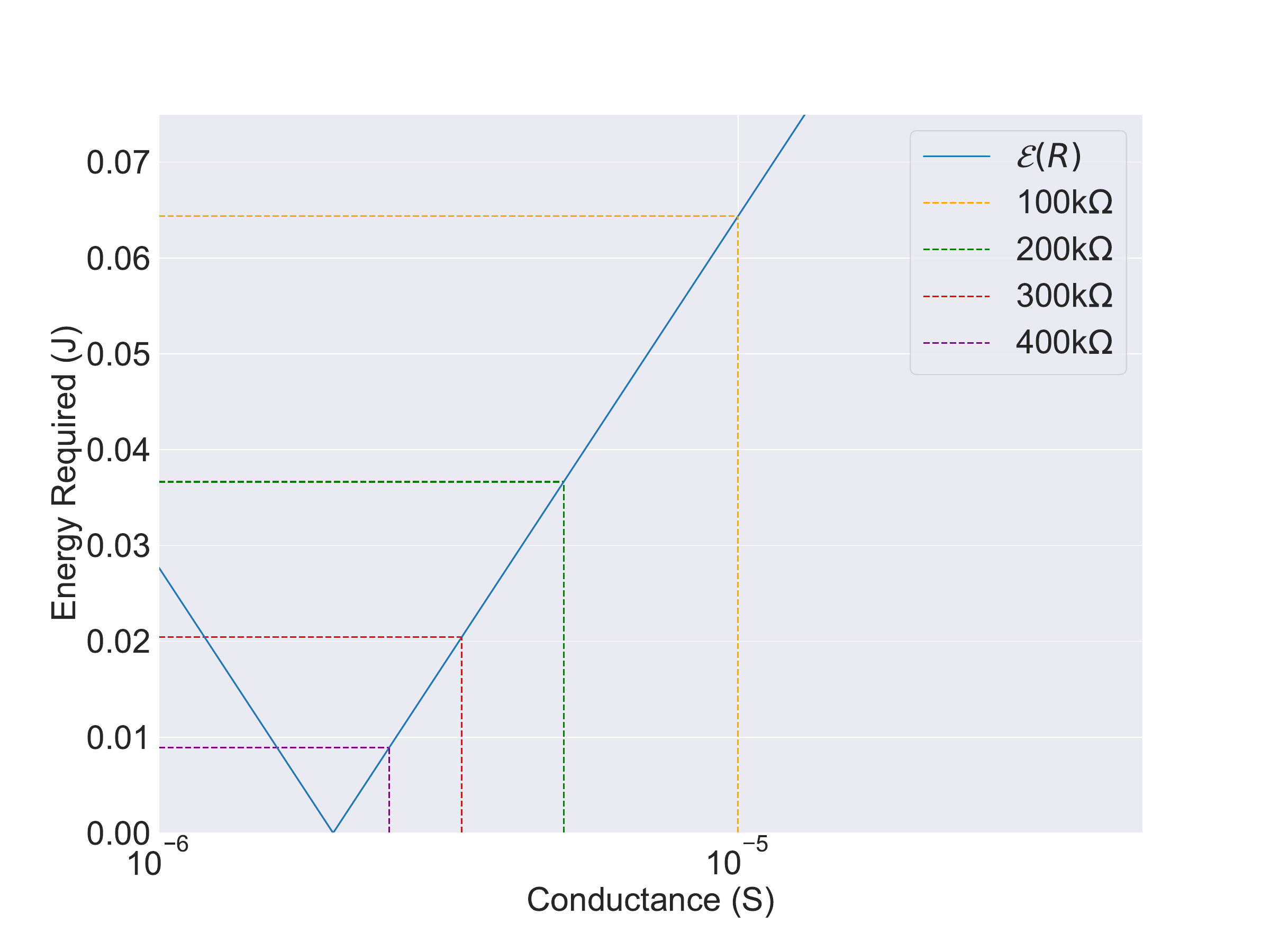}
    \caption{Costs associated with bringing the resistance of a memristor to a specified value, starting from \(R(0) = 500k\Omega\), for \(\tau_{\text{final}}=1.0 s\), \(K=2.0\), and \(R(\tau_{\text{final}})=0.1k\Omega\). In terms of resistance (left), and conductance (right).
    }
    \label{fig:energy_cost}
\end{figure*}

Then, we can write
\begin{align}
 \tau &= \tau_{\text{final}} \cdot\frac{\frac{1}{R(\tau)} - \frac{1}{R(0)}} { \frac{1}{R(\tau_{\text{final}})} - \frac{1}{R(0)}}
    \label{eq:time_as_func_resistance} \\
    &= \frac 1J \left( \frac{1}{R(\tau)} - \frac{1}{R(0)} \right),
\end{align}
where we have defined $J := \frac{1}{\tau_{\text{final}}}\left(\frac{1}{R(\tau_{\text{final}})} - \frac{1}{R(0)}\right)$.

Note that power is equal to \(i^2 R\) for a constant resistance, \(R\), and current, \(i\).
As noted above, the maximum (compliance) current is inversely proportional to the final resistance obtained after the full current pulse duration, \(\tau_{\text{final}}\), i.e.,

\begin{align}
    i_{\text{max}} &\propto \frac{K}{R(\tau_{\text{final}})}
    \label{eq:inverse_proportional_current_resistance}
\end{align}

In order to satisfy the boundary condition that \(i_{\text{max}} = 0\) for \(R(\tau_{\text{final}}) = R(0)\), we have:

\begin{align}
    i_{\text{max}} &= K \cdot \left(\frac{1}{R(\tau_{\text{final}})} - \frac{1}{R(0)}\right).
    \label{eq:inverse_relationship_corrected} 
 \end{align}

We will make the simplifying approximation that the current is always equal to its maximum compliance value during a current pulse (i.e., \(i(t) = i_{\text{max}}, \; \forall t \in [0, \tau_{\text{final}}]\)).
We can use Equations~(\ref{eq:resistance_as_func_time})~and~(\ref{eq:inverse_relationship_corrected}) to integrate the power (as a function of time) to obtain a relationship between the final resistance value and the energy expenditure, \(\mathcal{E}\), for a constant current and a final resistance value that is assumed inversely proportional to the duration of a current pulse of a fixed magnitude. We assume that the energy expenditure is symmetric; that is, the amount of energy required to change the state from \(R_{1} \rightarrow R_{2}\) is equal to that required to change it from \(R_{2} \rightarrow R_{1}\), thus taking the absolute value of the integral to obtain the energy expenditure:

\begin{align}
    \mathcal{E}(R(\tau)) \notag &= \left|\int_{t=0}^{t=\tau} i(t)^2R(t) dt \right| \\
    \notag &= \left|\int_{t=0}^{t=\tau} \left(K \cdot J \cdot \tau_{\text{final}}\right)^2 \frac{1}{\frac{1}{R(0)} + tJ} dt \right|\\
    \notag &= \left| \left[\left(K \cdot J \cdot \tau_{\text{final}}\right)^2 \frac{\ln \left|\frac{1}{R(0)}+ tJ\right|}{J} \right]^{\tau}_{0} \right| \\
    &= \left| K^2 \cdot J \cdot \tau^2_{\text{final}}  \ln{\left|1 + \tau J R(0)\right|} \right|
\end{align}

Substituting Equation~(\ref{eq:time_as_func_resistance}):

\begin{align}
    \mathcal{E}(R(\tau)) &= \left| K^2 \cdot J \cdot \tau^2_{\text{final}} \ln{\left|\frac{R(0)}{R(\tau)}\right|} \right|
    \label{eq:energy_resistance_relationship} \\
    & = \left| A \cdot \ln{\left|\frac{B}  {R(\tau)}\right|} \right|
\end{align}

Equation~(\ref{eq:energy_resistance_relationship}) has a choice of constant parameters \(A \triangleq \tau_{\text{final}} \cdot K^2 \cdot  \left(\frac{1}{R(\tau_{\text{final}})} - \frac{1}{R(0)}\right)\) and \(B \triangleq R(0)\), that respectively scale and determine the shape of the energy cost as a function of \(R\). In the next section, this energy cost function will be used as a regularisation term in the training of an autoencoder to store information as resistance values on memristors, resulting in a trade-off between the range of states available for storage (with those at lower resistance levels being more stable\footnote{Lower resistance states can be seen as channels that have a higher ``SNR'', due to the inverse relationship between the variance of the final resistance output for a given delay and the initial resistance value.})  and the average energy consumption required to bring the device down to that resistance level. In our experiments, we set \(R(0)\) equal to the equilibrium resistance, since this resistance can be achieved without expending any energy, through the autonomous drift of the resistance of the device, in the absence of a programming pulse.

The energy cost is shown in Figure~\ref{fig:energy_cost}, for \(\tau_{\text{final}}=1.0 s\), \(K=2.0\), \(R(0) = 500k\Omega\), and \(R(\tau_{\text{final}}) = 0.1k\Omega\). These are the parameters we use to define an illustrative energy cost for our experiments in Section~\ref{sec:autoencoder_storage}, where a linearly scaled energy cost is used as a regularisation in the optimisation objective (Equation~(\ref{eq:autoencoder_loss_function})).

\section{Method}
\label{sec:autoencoder_storage}
The delay from the initial storage of information to recovery in the presence of resistive drift is analogous to the noise power in the classical \ac{AWGN} channel, where the transmission power can be altered to improve the \ac{SNR}.
In order to train an end-to-end \ac{DeepJSCC} scheme \cite{bourtsoulatzeDeepJointSourceChannel2019}, we require a differentiable channel model. To this end, we make use of a generative modelling approach for resistive drift in memristors, which uses a \ac{cGAN} as a differentiable and computationally simple channel model \cite{el-geresyDelayConditionedGenerative2024}. We use this to model the delay and initial resistance conditioned resistive drift channel distribution.

\subsection{Dataset}

We use a resistive drift dataset based on an event-based model of memristive devices \cite{el-geresyEventBasedSimulationStochastic2024}, introduced in \cite{el-geresyDelayConditionedGenerative2024}, where a differentiable \ac{cGAN} channel model is trained to approximate the statistics of the simulated resistive drift using a data-driven approach. The proposed generated time-series dataset consists of 5000 series, each of length 1000s with resistance values uniformly sampled at a rate of 1Hz (i.e., 1001 points per series), with initial resistance conditions spaced uniformly over the range \([100\Omega, 750k\Omega]\).

\subsection{Data Normalisation}

Data normalisation forms an essential initial step for ensuring our training data is presented to the models in a suitable form.
In the case of the resistance values, we make use of the same resistance normalisation transform as in \cite{el-geresyDelayConditionedGenerative2024}. We do not explicitly make use of this transform during the training procedure. The trained \ac{cGAN} model assumes input and output resistances to be in the normalised space. Thus, we simply allow our \ac{DeepJSCC} autoencoder to generate values that are assumed to already be normalised resistance values.

We here re-iterate the data transform introduced in \cite{el-geresyDelayConditionedGenerative2024}, which consists of a logarithmic transform followed by subtraction of the mean and scaling by the standard deviation (using empirical values averaged over the drift dataset) in order to normalise the magnitudes. Let \(\mu_R\) denote the sample mean of the logarithmically transformed resistances and \(\sigma_R\) the square root of the sample variance of the resistances:

\begin{align}
    \notag
    \mu_R &:= \frac{1}{|\mathcal{D}|}\sum_{S_i \in \mathcal{D}}\frac{1}{|S_i|}\sum_{r_j^i \in S_i} \ln{(r_j^i)}, \\
    \sigma_R &:= \sqrt{ \frac{1}{|\mathcal{D}|}\sum_{S_i \in \mathcal{D}}\frac{1}{|S_i|}\sum_{r_j^i \in S_i} (\ln{(r_j^i)} - \mu_R) },
    \label{eq:dataset_std}
\end{align}
where \(\mathcal{D}\) denotes the whole (data)set of time series, with \(S_i\) being the \(i\)'th time series in the dataset, and \(r_j^i \in S_i\) the ordered (resistance) elements of each series, indexed by \(j\). Thus, the resistance normalisation transformation, \(N_{\text{res}}(r)\), is given as:

\begin{align}
    N_{\text{res}}(r) = \frac{\ln{(r)}-\mu_R}{\sigma_R}.
    \label{eq:data_transform}
\end{align}

We thus pass the outputs of \(E_{\theta_e}\) - the encoder - untransformed to the generator model of the trained \ac{cGAN} channel model. Similarly, the input to \(D_{\theta_d}\) - the decoder - is assumed to be in the normalised resistance space.

The autoencoder's goal is to transform the input distribution contingent upon the delay, and there may be a highly non-linear relationship between the input delay and the associated transformation. Thus, we apply a logarithmic transformation to the delay before passing it to the encoder and/or decoder networks, as shown in Equation~(\ref{eq:autoencoder_delay_transform}). In this transform, we add a small constant to the delay before taking the logarithm, to lower bound the output and prevent it from reaching \(-\infty\). Note that in the case of the autoencoder, only positive delays are considered, thus \(T \geq 0\), ensuring a valid domain. The transformation is dependent on the delays we use for training, which are uniformly distributed integers in the range \([t_{\text{min}}, t_{\text{max}}]\). In the case that \(T\) has support of only a single possible value (i.e. it is a degenerate distribution and \(t_{\text{min}} = t_{\text{max}}\)), this constitutes training for a single delay condition\footnote{This scenario never arises in delay-conditioned experiments, since if there is only one delay condition, then conditioning on the delay is redundant; however, we include this point for the sake of completeness.} and we set \(\mu_T = 0\) and \(\sigma_T = 1\), otherwise, we compute the (logarithmic) delay normalisation transform for the autoencoder according to:

\begin{align}
    \notag
    \mu_T &= \mathbb{E}_T[\ln{(T + 0.1)}] \\
    \notag
    \sigma_T &= E[(\ln{(T + 0.1)} - \mu_T)^2] \\
    \bar{t} &= \frac{\ln{(t + 0.1)} - \mu_T}{\sigma_T}
    \label{eq:autoencoder_delay_transform}
\end{align}

\begin{figure}
    \centering
    \includegraphics[width=\linewidth]{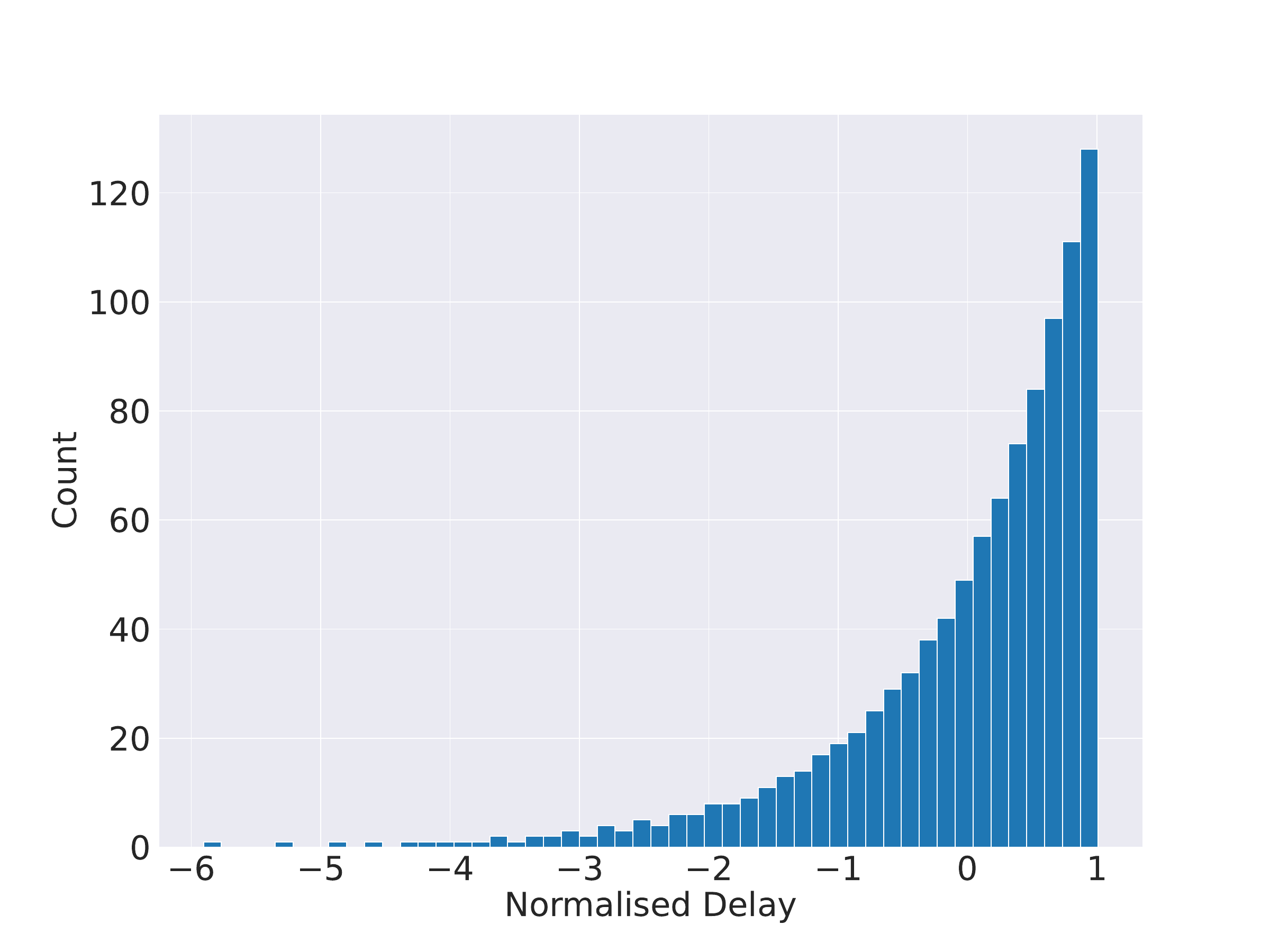}
    \caption{Histogram of the normalised delays following the logarithmic transformation used for training and conditioning the autoencoder.
    }
    \label{fig:hist_delay_transform}
\end{figure}

A histogram of the transformed delays in the dataset (delays in the range \([1, 1000]\)) is visualised in Figure~\ref{fig:hist_delay_transform}.

\subsection{Regularisation}

We introduce regularisation terms to satisfy the energy constraints based on the derived energy cost function for a particular input distribution (storage scheme), as well as constraints on the range of allowable encoded resistance values. We use two forms of regularisation for training our network, transforming the problem from simple optimisation of the \ac{MSE} of the reconstructed image into one where a trade-off must be made between three loss terms: the \ac{MSE}, a resistance regularisation term, and an energy regularisation term.

\subsubsection{Resistance Regularisation}

The first regularisation term is used to ensure that the resistances output by the encoder network fall within a valid range. Note that the logarithmic transform applied to the resistances, as given in Equation~(\ref{eq:data_transform}), means that we take an exponential transform (the inverse) at the output of the encoder, ensuring that the encoded resistance values are always positive.

In order to enforce these limits, we use a squared error loss when the values at the encoder output violate the boundaries \cite{zarconeJointSourcechannelCoding2018}.
We make use of hard regularisation limits in addition to the soft limits, to prevent pathological encoder outputs early on during the training process, to avoid the problem of gradient explosion, where excessively large gradient values during optimisation may cause training to fail early on.

We denote the upper, soft regularisation limit by \(R_{\text{high}} \leq 750 k\Omega\). Similarly, we denote the lower, soft regularisation limit by \(R_{\text{low}} \geq 0.1k\Omega\). The hard upper and lower limits are denoted by \(b_{\text{high}}\) and \(b_{\text{low}}\), respectively, and set to be equal to the maximum and minimum resistance values we want our network to be able to generate, in our case, \(100 M\Omega\) and \(1 \Omega\), respectively. The soft limits are enforced through regularisation, whereas the hard limits are enforced through non-differentiable encoder output clipping at the given values.
We compute the soft regularisation loss in the transformed logarithmic space, to allow for fine-grained control that is less dependent on order of magnitude.

The upper soft resistance regularisation term, \(r_{\text{upper}}(\cdot)\), is given as follows for the output of the encoder, \(M\):
\begin{align}
    r_{\text{upper}}(X) = \frac{1}{n}\sum_{i} (\max(M_i-R_{\text{high}}, 0))^2
    \label{eq:upper_resistance_regulariser}
\end{align}
and the lower soft resistance regularisation term, \(r_{\text{lower}}(\cdot)\), is given as:
\begin{align}
    r_{\text{lower}}(X) = \frac{1}{n}\sum_{i} (\max(R_{\text{low}}-M_i, 0))^2
    \label{eq:lower_resistance_regulariser}
\end{align}

\begin{figure*}
    \centering
    \includegraphics[width=\linewidth]{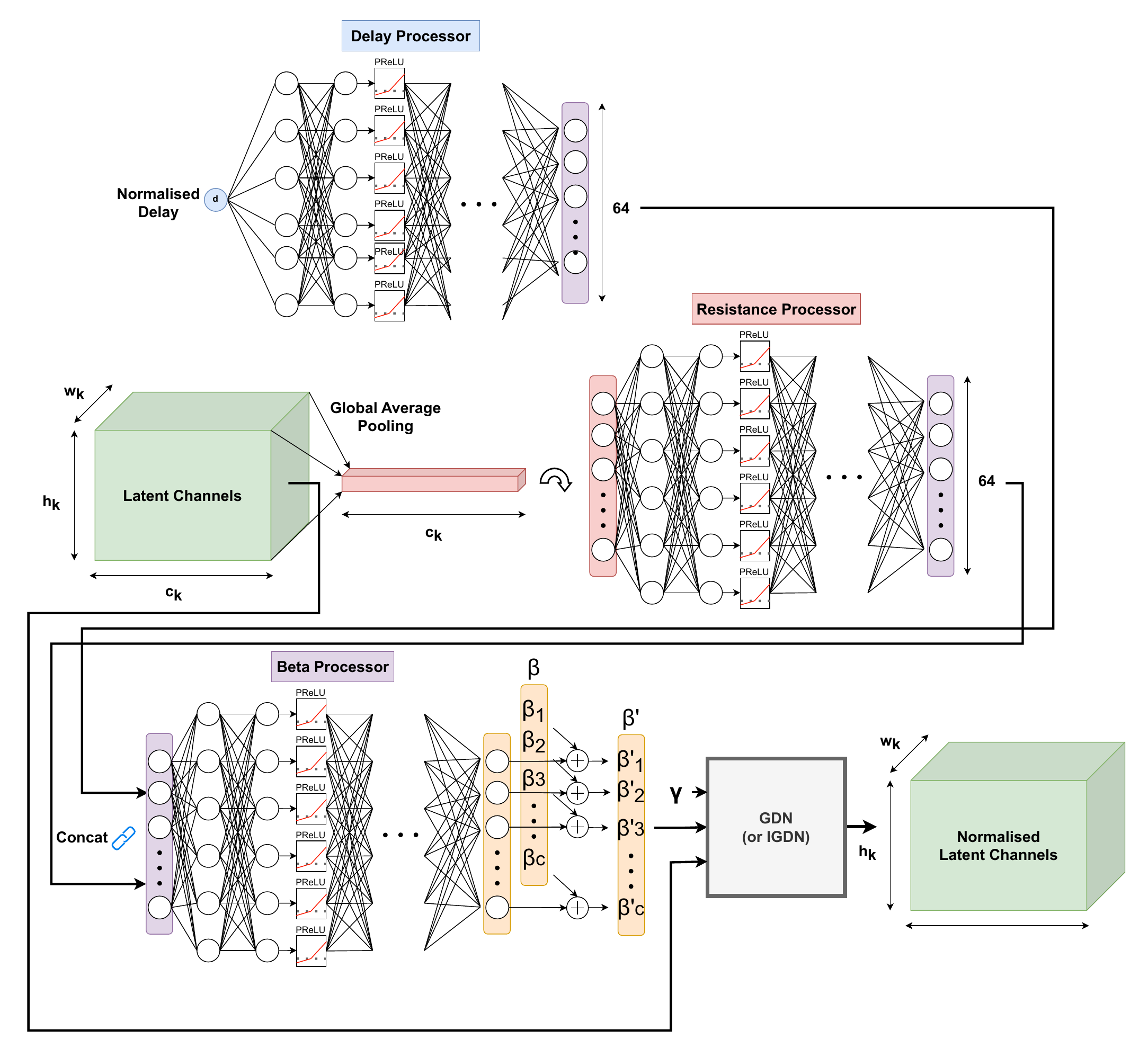}
    \caption{The \ac{cGDN} transform for conditioning the encoder and/or decoder network on the delay, applied to the latent channels outputs for each convolutional layer for the encoder/decoder conditioned settings.}
    \label{fig:cgdn}
\end{figure*}

\subsubsection{Energy Regularisation}

The second regularisation term is based on the energy cost function in Equation~(\ref{eq:energy_resistance_relationship}) and forms the basis of the trade-off between the use of stable resistance states, which are less prone to resistive drift (and information decay) for a given delay, and the image reconstruction error. We specify an energy budget \(e_b\), which is the upper limit on the average energy (the power) permitted for use in encoding information. The energy regularisation function \(\mathcal{E}(\cdot)\) was given in Equation~(\ref{eq:energy_resistance_relationship}). Making use of \(\mathcal{E}(\cdot)\), given its necessary parameters, our second regularisation term, the energy regularisation term \(r_{\text{energy}}\), is given as:
\begin{align}
    r_{\text{energy}}(M) = \frac{1}{n}\sum_{i} \left(1 - \mathcal{E}\left(\frac{M_i}{e_b}\right)\right)^2,
    \label{eq:energy_regularisation_term}
\end{align}
where \(X\) is the image to be encoded and \(\mathcal{E}\) is our energy regularisation function. Note that we normalise the output energies by the budget, \(e_b\), to achieve a consistent scaling, regardless of the absolute magnitude of the energy budget, in the energy regularisation.

Thus, we obtain our loss function to be optimised, shown in Equation~(\ref{eq:autoencoder_loss_function}), with parameters \(\lambda_{\text{res}}\) and \(\lambda_{\text{energy}}\) being design choices used to adjust the stringency of regularisation relative to the main reconstruction error. In our experiments, we set \(\lambda_{\text{resistance}} = \lambda_{\text{energy}} = 1\). Let there be random sets composed of \(b\) samples of \(X\) - mini-batches - and let the indexed elements of a given mini-batch be denoted by \(X_i,  \forall i \in [1, b]\).

\begin{strip}
\begin{align}
    l(X, \hat{X})= \sqrt{\sum_{i=1}^{b}(X_i-D_{\theta_d}(\hat{M}(T, E_{\theta_e}(X))))^2)} + \lambda_{\text{resistance}} \left( \sum_{i=1}^{b}r_{\text{upper}}(M) + \sum_{i=1}^{b}r_{\text{lower}}(M) \right) + \lambda_{\text{energy}} \left( \sum_{i=1}^{b}r_{\text{energy}} (M) \right).
    \label{eq:autoencoder_loss_function}
\end{align}
\end{strip}

\begin{figure*}
\centering
\includegraphics[width=\linewidth]{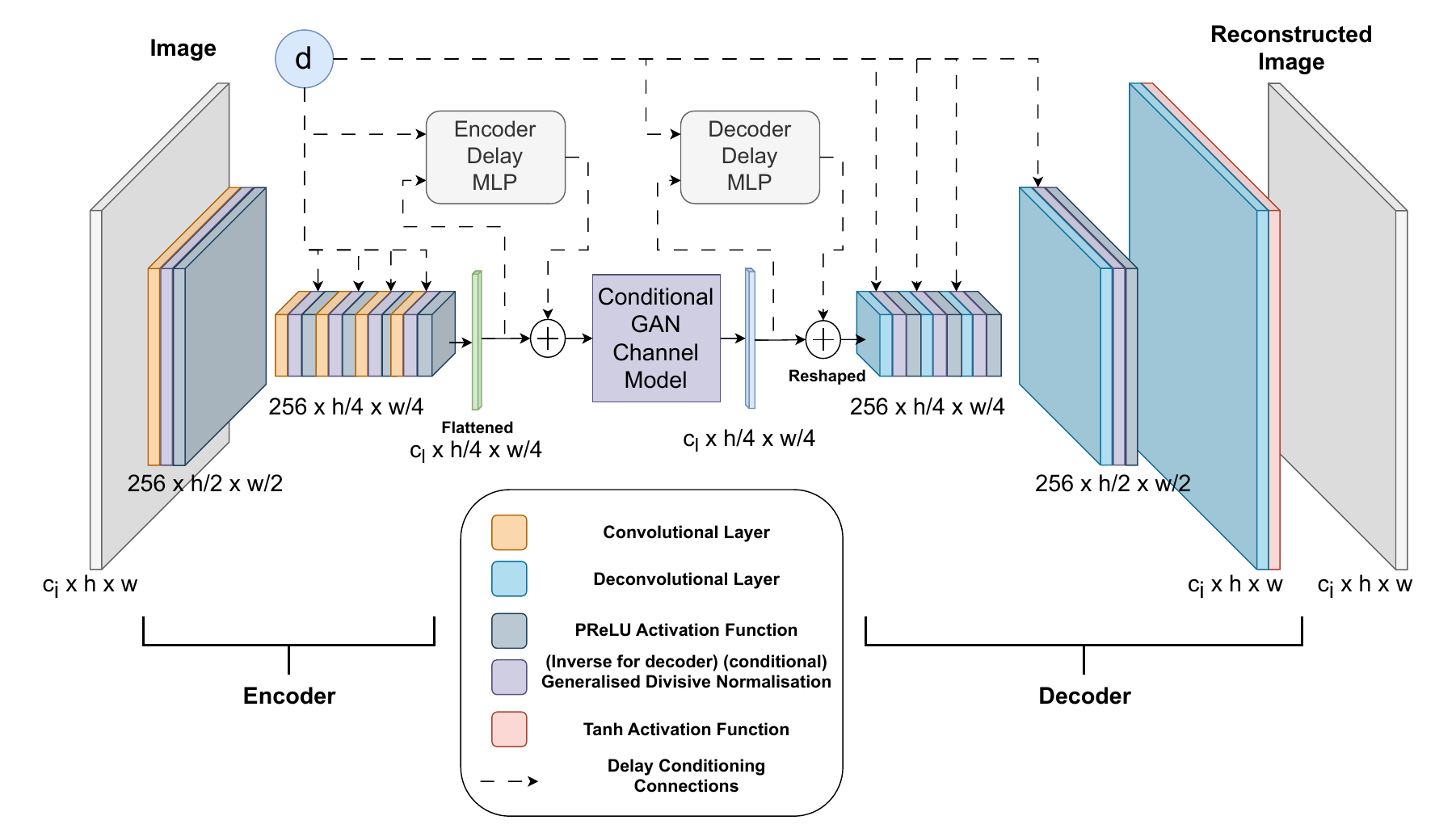}
\caption{The \ac{DeepJSCC} autoencoder architecture for delay-conditioned data storage. To introduce delay conditioning, we replace the \ac{GDN} layers in \cite{bourtsoulatzeDeepJointSourceChannel2019} with \ac{cGDN} layers, as well as introducing fully connected residual delay processors at the output of the encoder and/or the input of the decoder. \ac{cGDN} and the delay processors are only included for the encoder/decoder, respectively, if the setting includes delay conditioning. In the diagram, \(c_i\) are the number of colour channels in the image, and \(c_l\) are the number of latent channels. Changing \(c_l\) modifies the compression rate of the autoencoder.}
\label{fig:kurka_networks}
\end{figure*}

\subsection{Conditional Generalised Divisive Normalisation}

\Acf{GDN} is a form of normalisation designed to mimic a canonical computational primitive in the human brain \cite{burgLearningDivisiveNormalization2021}, namely the phenomenon of divisive normalisation.
A form of \ac{GDN} has been used in image compression to improve the learning of an informative representation for source coding of images \cite{balleEndendOptimizedImage2017}. It is an alternative to other kinds of normalisation, such as batch normalisation and layer normalisation. \ac{GDN} has been used in \ac{DeepJSCC} autoencoder architectures \cite{bourtsoulatzeDeepJointSourceChannel2019} for the transmission of natural images. For a given convolutional output layer index \(k\), let us denote the value of the \(i\)'th channel at spatial location \((m,n)\) by \(w_i^{(k)}(m, n)\). The output of the \ac{GDN} transform for this weight, denoted by \(u_i^{(k)}(m,n)\), is defined as follows \cite{balleEndendOptimizedImage2017}:
\begin{align}
    u_i^{(k)}(m, n) = \frac{w_i^{(k)}(m, n)}{\sqrt{\beta_{k, i} + \sum_j{\lambda_{k, ij}(w_j^{(k)}(m, n))^2)}}},
    \label{eq:gdn}
\end{align}
where the set of parameters \(\beta_{k, i}\) - computed per channel - and \(\gamma_{k, ij}\) - modulating the influence of weights at the same spatial location in different channels on each other - are to be learnt. Note that the normalisation uses the square of each weight \(w_j^{(k)}(m, n)\), ensuring that the magnitude, rather than the sign, is used to normalise the neighbouring weights across the channel dimension.

We can consider delay information to be a separate modality from the image to be compressed. We take inspiration from a conditional adaptation of \ac{BN}, which is known as \ac{cBN} \cite{devriesModulatingEarlyVisual2017}. This technique was originally introduced to allow for the modulation of the processing of visual information (images) using linguistic information (language) to allow for the learning of better representations for \ac{VQA} problems. We adapt this approach to introduce a conditioning to \ac{GDN} and we term it \acf{cGDN}.

The \ac{cGDN} transform is shown in Figure~\ref{fig:cgdn}, and is applied to the output of each convolutional layer. We use a fully connected processor with non-linear activations to form an embedding of the delay. Simultaneously, using global average pooling to reduce the dimensionality, along with a fully connected processor, we form an embedding for the activations of the output of the convolutional layer. Both representations are fed to a fully connected combined beta processor network to produce a vector of dimension \(\mathbb{R}^{c_k}\), where \(c_k\) is the number of channels in layer \(k\) of the convolutional architecture of the encoder. The output vector is then added to the \(\beta\) parameters for that layer,
which are indexed by \(i\) in Equation~\ref{eq:gdn}.

The corresponding inverse transformation (\ac{icGDN}), that is used at the output of each of the decoder devconvolutional layers, is conditioned in the same way. Note that weights are \textit{not} shared between the corresponding forward and inverse \ac{cGDN} transforms.

\subsection{Additional Residual Delay Conditioning}

To condition the encoder and/or decoder on the delay, we also include an additional fully connected residual delay processor, that takes \(d\) along with either \(R\) or \(\hat{R}\) (i.e., the latent representation at the output of the encoder, or the input to the decoder) and transforms them into a residual vector of dimension \(\mathbb{R}^{c_l \cdot h_l \cdot w_l}\), which is added pointwise to the given resistance input.

\subsection{DeepJSCC Architecture}

The architecture of the autoencoder is shown in Figure~\ref{fig:kurka_networks}. This architecture is based on an existing deep convolutional \ac{DeepJSCC} architecture \cite{bourtsoulatzeDeepJointSourceChannel2019}, with \ac{GDN} modules replaced with a \ac{cGDN} module, which are passed the delay in order to condition the encoder and/or decoder transform(s) on the delay condition, as well as the inclusion of residual fully connected delay processor layers at the output of the encoder/input to the decoder, if the encoder/decoder are delay conditioned.

\subsection{Delay Specialisation}
\label{sec:delay_specialisation}

We find that training on batches with a \textit{single delay} condition at a time improves the ability of the autoencoder network to tailor its function to the delay condition. As such, we enforce the energy constraint using a separate set of (fixed) delays, belonging to the set \(\mathcal{D}_{\text{energy}}\), which is defined for a particular batch size \(b\), with delays for a uniform delay distribution in the range \([d_{\text{min}}, d_{\text{max}}]\):

\begin{align}
    \notag a &\triangleq \frac{d_{\text{max}} - d_{\text{min}}}{b - 1}, \\
    \mathcal{D}_{\text{energy}} &= \{d_{\text{min}}, d_{\text{min}} + a, \ldots, d_{\text{max}}\}.
\end{align}

Given each batch of the encoder, we compute the energy regularisation term separately to the other losses, according to the delays \(d \in \mathcal{D}_{\text{energy}}\). This ensures that the energy constraint need not be perfectly satisfied for any given random batch of delays sampled to compute the reconstruction error, but must only be obeyed on average according to the uniform delay distribution, which allows the autoencoder to learn to trade off the energy budget among different delay conditions.

\subsection{Training Parameters}

In \cite{el-geresyDelayConditionedGenerative2024}, it was shown that the \ac{cGAN} model trained with a delay discriminator could reasonably match the statistics not only within the bounds of the delays present in the training data, but also up to the bounds of the delays supplied during delay discrimination (i.e. \([1, 500]\)). In order to train the autoencoder for delays larger than this maximum value, we choose to evaluate the \ac{cGAN} recurrently \(n\) times, extending the validity of the delay conditions to \(500n\). We trade off the value of \(n\) for the computational complexity and risk of vanishing gradients when evaluating the \ac{cGAN} recurrently, choosing a relatively conservative value of \(n=4\), ensuring that the model is valid for delays up to 2000, which is within the range of delays we choose to train for: \(d \in [0, 1000]\).

We set the batch size \(b = 32\) for training. We use Adam optimiser \cite{kingmaAdamMethodStochastic2017} with a learning rate of \(5 \times 10^{-5}\). We train each delay conditioned scenario (see below) for 50 epochs on the CIFAR-10 dataset \cite{krizhevskyLearningMultipleLayers2009}, optimising the \ac{DeepJSCC} autoencoder to minimise the \ac{MSE}.

\section{Results}
\begin{figure*}
    \centering
    \hspace*{\fill}
    \subfloat[Energy budget 1.0J.]{
      \includegraphics[width=0.3\linewidth]{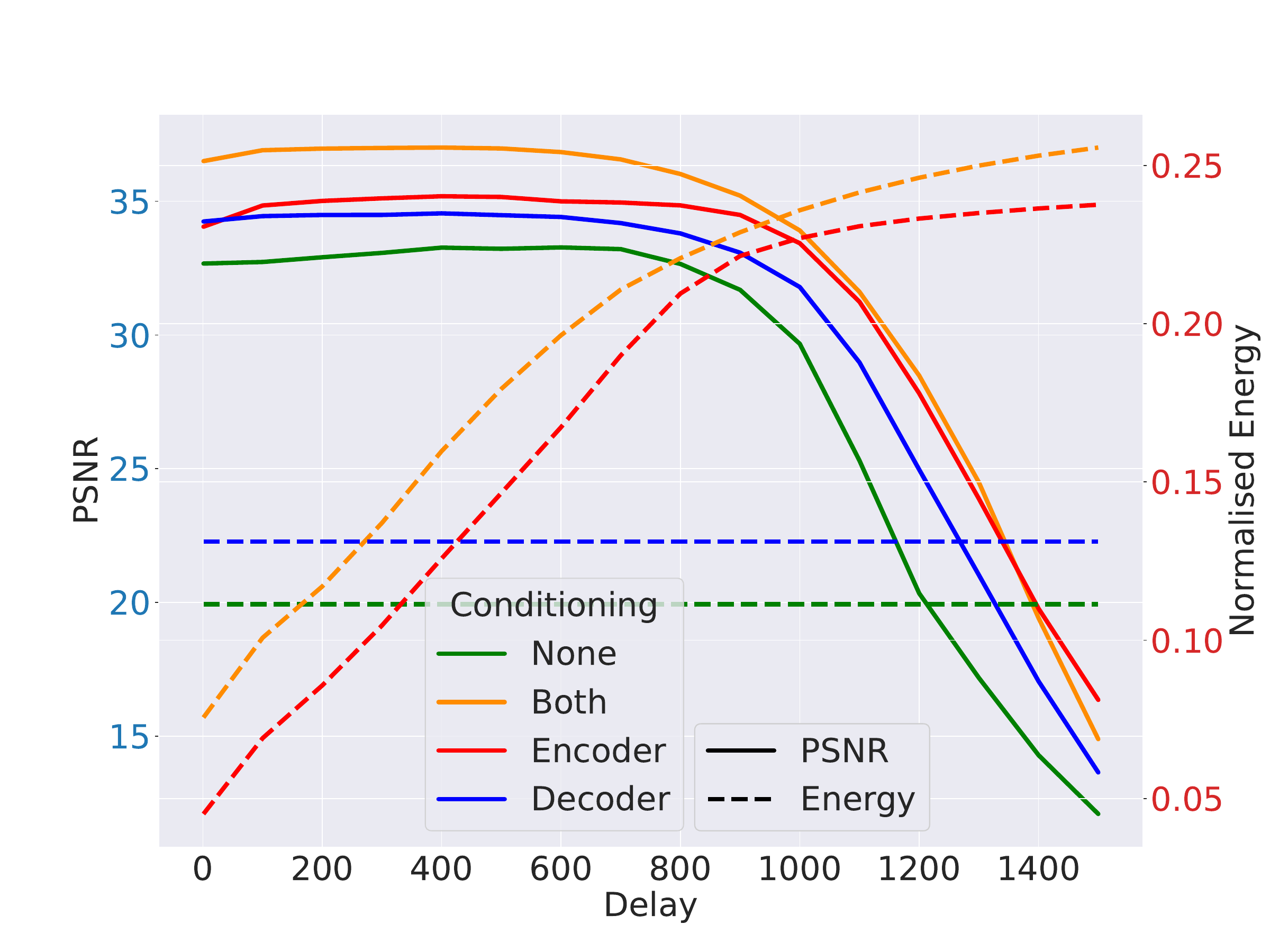}
      \label{fig:psnr_1.0}
    }
    \hfill
    \subfloat[Energy budget 0.5J.]{
      \includegraphics[width=0.3\linewidth]{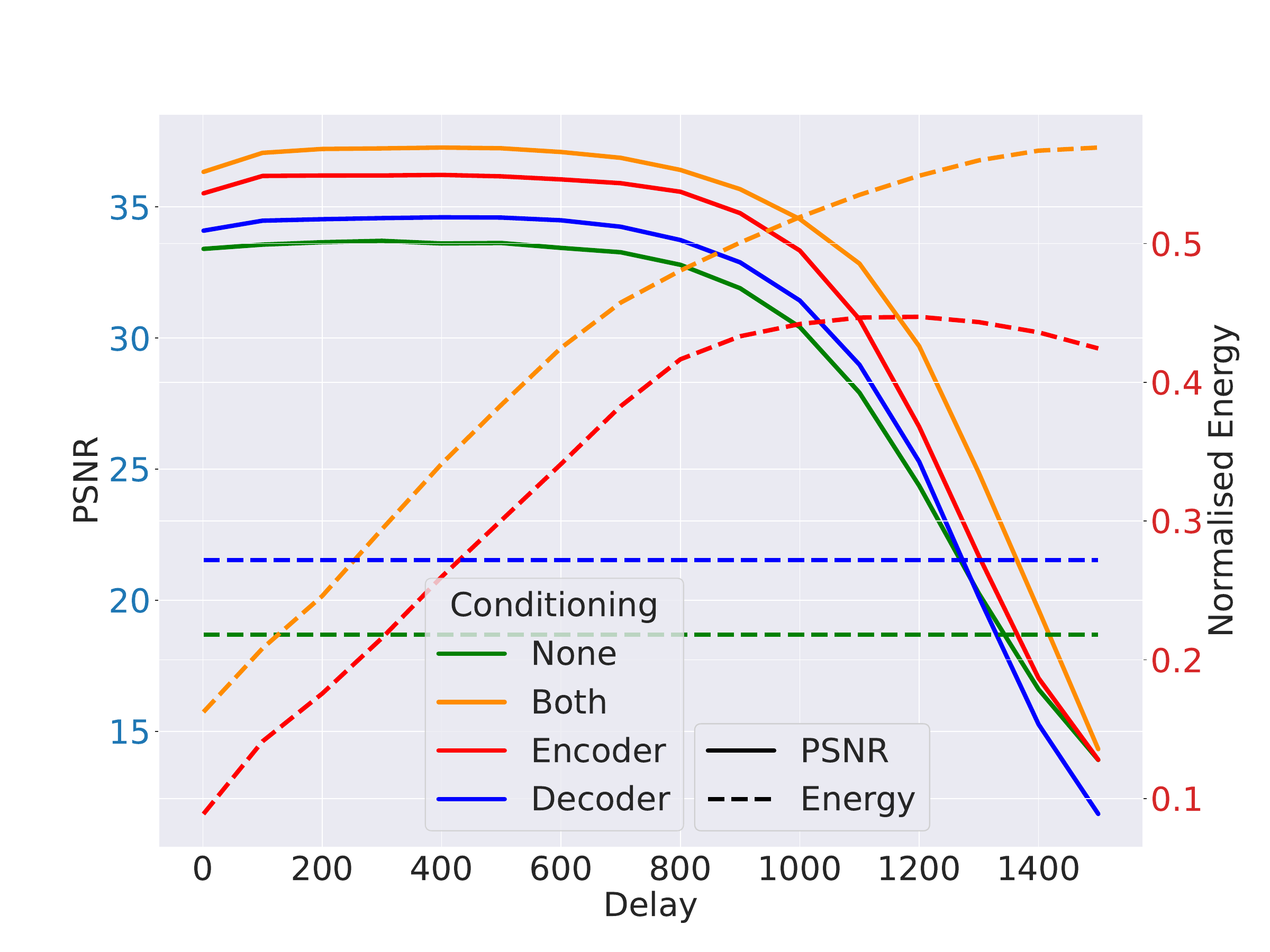}
      \label{fig:psnr_0.5}}
      \hfill
    \subfloat[Energy budget 0.1J.]{
      \includegraphics[width=0.3\linewidth]{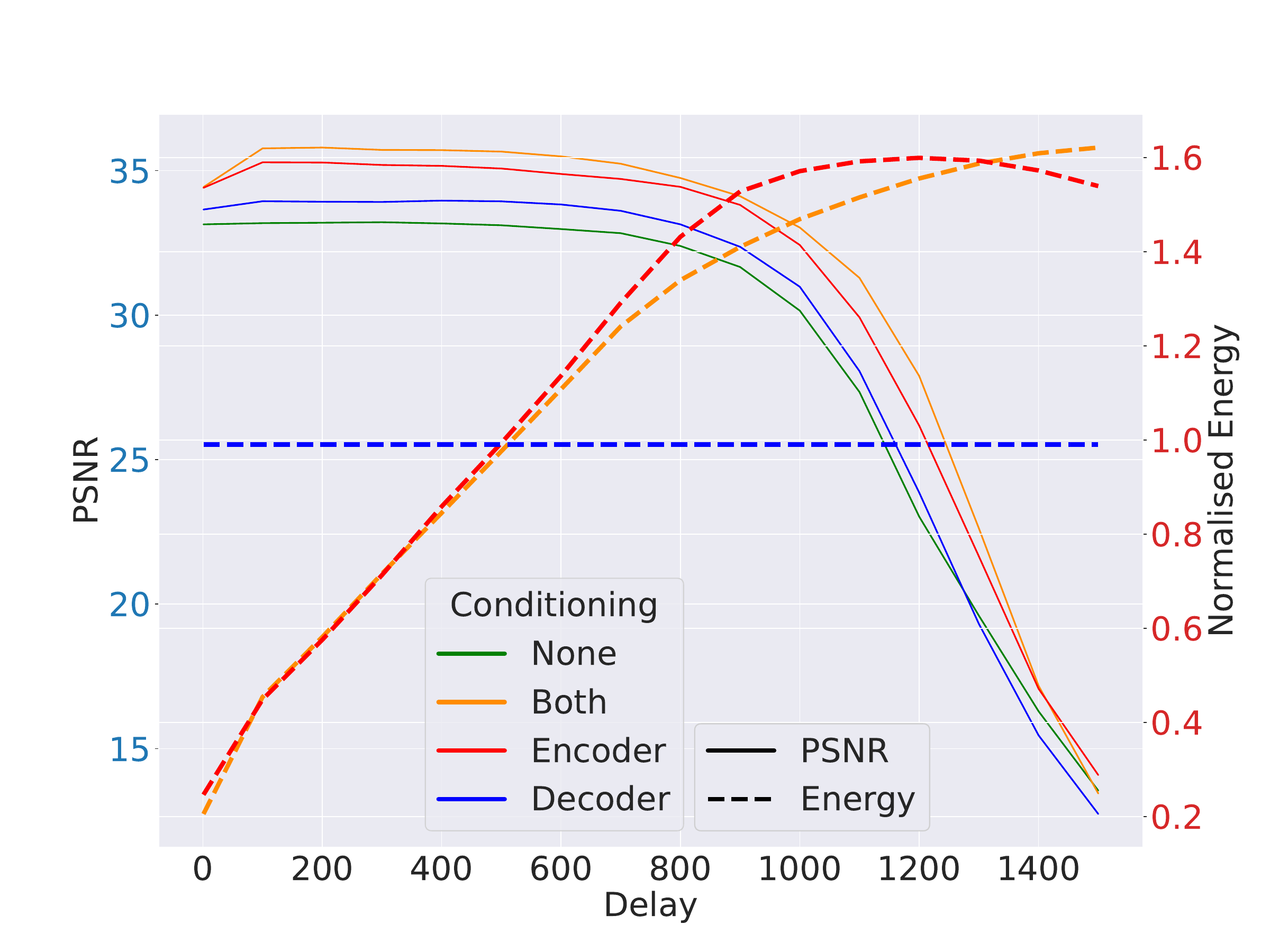}
      \label{fig:psnr_0.1}}
      \hspace*{\fill}
      \\
      \hspace*{\fill}
    \subfloat[Energy budget 0.05J.]{
      \includegraphics[width=0.3\linewidth]{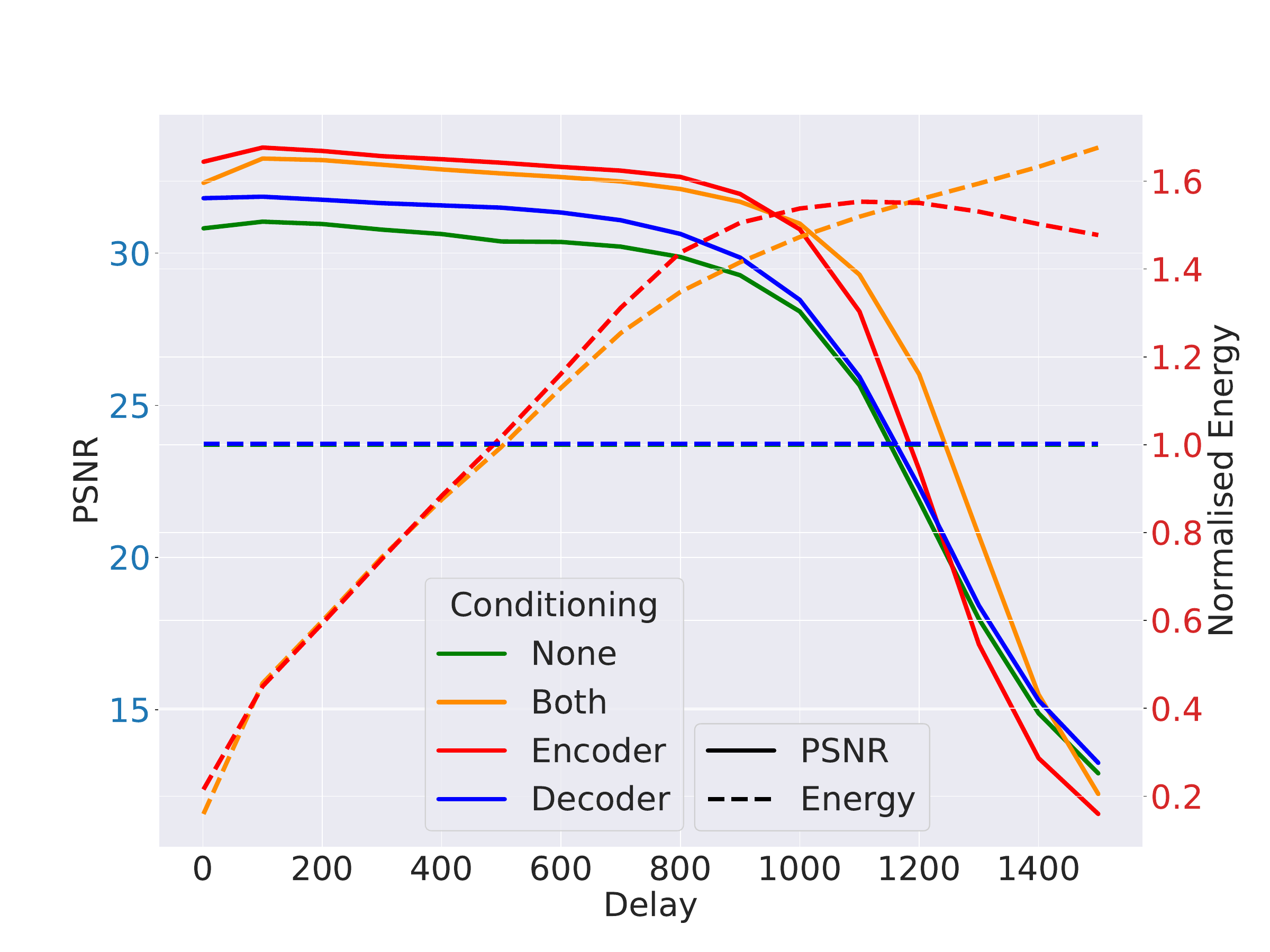}
      \label{fig:psnr_0.05}}
      \hfill
    \subfloat[Energy budget 0.01J]{
      \includegraphics[width=0.3\linewidth]{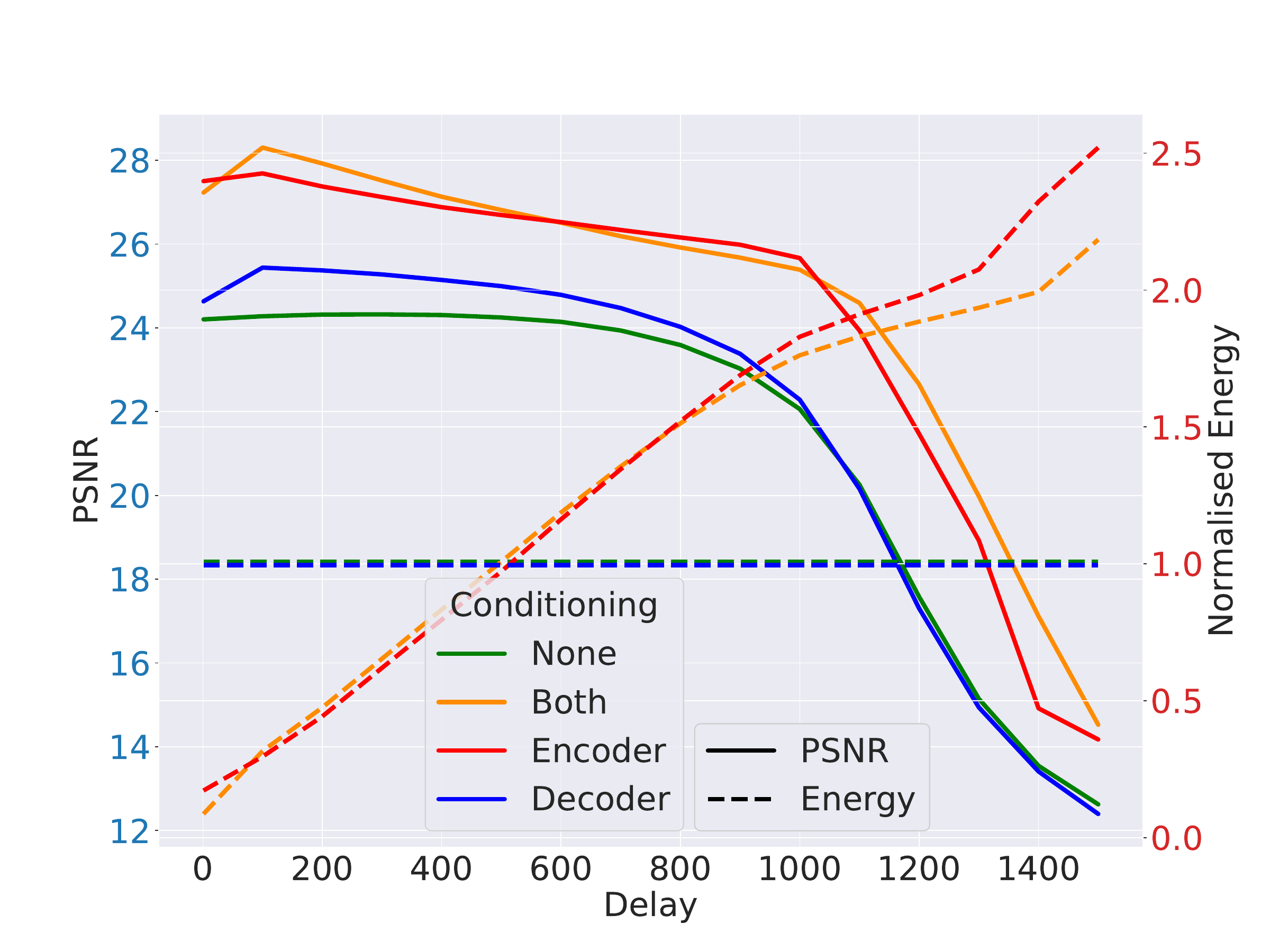}
      \label{fig:psnr_0.01}}
      \hspace*{\fill}
    \caption{A comparison of the \acp{PSNR} for a variety of autoencoder experiments, demonstrating the impact of delay conditioning of the encoder/decoder networks. Introducing delay conditioning to the encoder, allows the autoencoder to allocate different energy budgets to different conditions, while still respecting the average energy budget. Decoder conditioning also results in improved results compared to the absence of conditioning, with the decoder able to correct for delay-dependent distortions, resulting in better SNR reconstruction values across a wider range of the delays considered.}
    \label{fig:autoencoder_training_results}
\end{figure*}

We train the autoencoder using four different experimental settings. The task is to reconstruct image from the stored resistance values after a given storage delay. The network learns to minimise the error for the given delay distribution, learning the best distribution of resistances to program that will result in a minimum expected distortion.

\begin{enumerate}
    \item In the ``unconditioned'' setting, we train the encoder and decoder without knowledge of the delay. Thus, the autoencoder must learn the best ``average'' encoding and decoding scheme that minimises the error over the given range of delay conditions.
    \item In the ``encoder conditioned" setting, the encoder is passed information about the delay, through the use of the aforementioned \ac{cGDN} modules and the encoder fully connected residual network. In this setting, the encoder must apply a pre-transform to the input distribution such that using a single decoding strategy, the \ac{MSE} is minimised. This scenario, though unrealistic, is shown to isolate the performance benefits offered by conditioning the encoder alone on the delay.
    \item In the ``decoder conditioned'' case, the decoder knows the delay, while the encoder is delay-agnostic. Hence, the decoder can use the delay information to better decode the stored image. In this setting, the encoder stores the image in the same manner no matter when it will be recovered, so its goal is to allow for decoding at the best possible quality, regardless of the delay.
    \item Finally, in the ``both conditioned'' (or simply ``conditioned'') setting, both the encoder and decoder networks are aware of the delay between storage and reading. Note that this scheme is more practical, compared to training a specific encoder/decoder pair for individual delays, since the single encoder/decoder pair that is trained can be used to store and recover information for any delay.
\end{enumerate}

\subsection{Reconstruction Results}

Figure~\ref{fig:autoencoder_training_results} shows the reconstruction performance (average \ac{PSNR}) for a variety of delay conditions, for each of the conditioning settings. We can see from the figure that for energy budgets \(e_b \geq 0.5J\), the networks are able to learn effectively using less than the required energy budget, however the delay conditioning of the networks still benefits training. The disparity in the energy budgets stems from the encoder making use of different encodings, dependent on the delay, however energy is not a bottleneck.
We see that as the budget falls to 0.1J, energy is maximised, making it clear that the energy budget begins to become a limiting factor for the performance. For 0.05J and 0.01J, we see steady decreases in the average \ac{PSNR}, with the delay-conditioned networks consistently outperforming the unconditioned version. In the case of decoder conditioning, we see that the network is able to better adjust for the deterministic distortions introduced by the channel, improving the performance; however, we observe that the introduction of encoder conditioning is of greater benefit, in terms of the reconstruction error, while conditioning of both the encoder and the decoder produces a synergistic effect that results in a higher performance than is possible with each of the individual networks, in most cases.

The encoder network is able to generalise to delays outside of the distribution given. Energy constraints are violated in order to ensure that enough information is retained, following storage and recovery, to reconstruct the image to a reasonable degree. The violation of the energy constraints demonstrates that the trade-off that the network has learnt between energy consumption and storage reliability generalises to larger delays.

\begin{figure*}
    \centering
    \includegraphics[width=\linewidth]{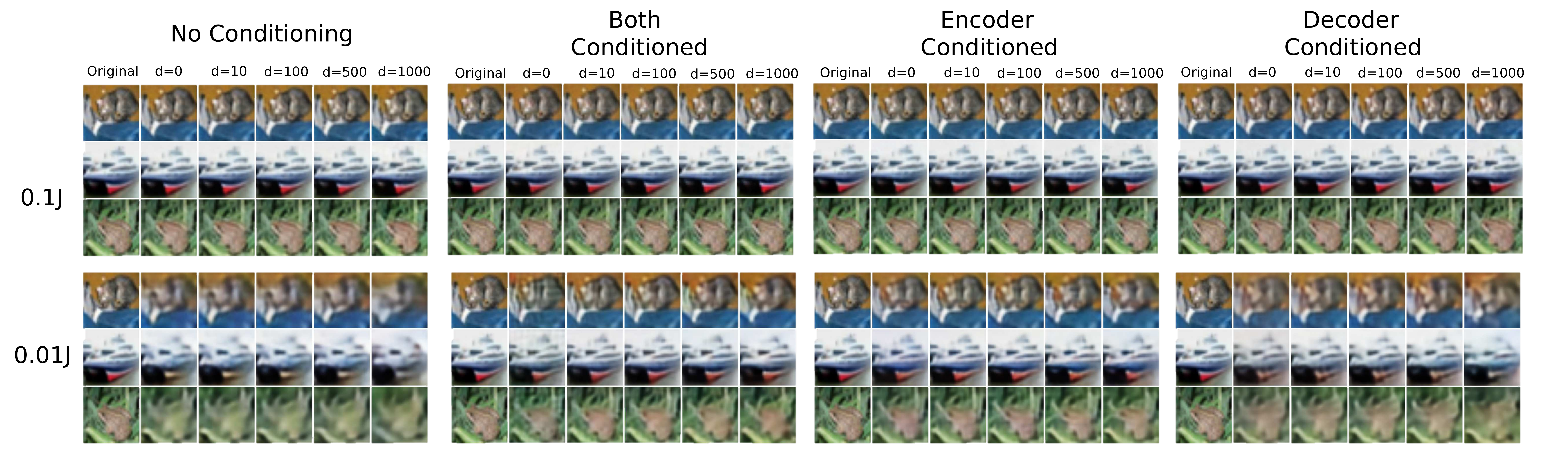}
    \caption{Image reconstruction results for a range of delays ranging from 0s to 1000s, for two different  energy budgets.}
    \label{fig:image_reconstructions}
\end{figure*}

Figure~\ref{fig:image_reconstructions} shows image reconstruction results under a range of channel delays, for the different delay-conditioned settings studied, for two different energy budgets (0.1J and 0.01J).  see that reconstruction quality generally improves with decreasing delay, except in the case of the unconditioned network, which uses a single encoding and decoding scheme, tailored to the noisiest setting. Therefore, particularly for very low energy budgets, conditioning is essential for achieving a reasonable reconstruction quality.

\subsection{Single Delay and Noiseless Ablation Experiments}

\begin{figure}
    \centering
    \includegraphics[width=\linewidth]{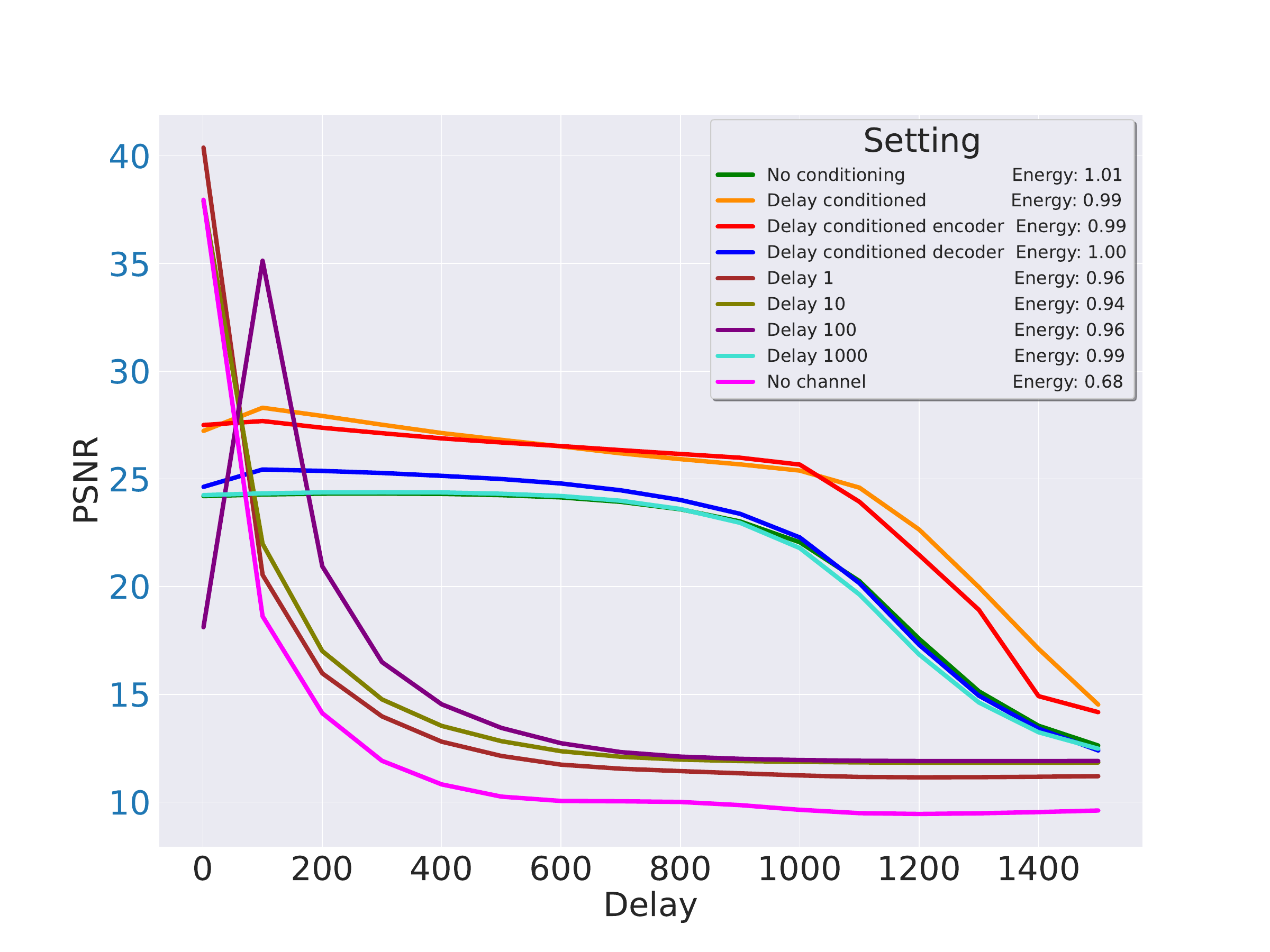}
    \caption{Comparison of the encoder/decoder/encoder and decoder -conditioned settings for autoencoder training to a range of ablation experiments, for training, without delay conditioning, on single delays only (1, 10, 100, 1000) and on a noiseless channel. We see that although training for a particular channel setting (delay) can improve the performance for that particular condition, the performance is much worse across other delays. The legend displays the average energy consumption for each of the scenarios. Networks specialised for only a single delay do tend to outperform the delay conditioned versions for the delay conditions for which they are specialised, at the expense of a drastically poorer performance across other delays.
    }
    \label{fig:single_delay_ablation}
\end{figure}

We compare the performance of the delay conditioned experiments for an energy budget of 0.01J to the performance of autoencoders trained for specific delay conditions, as well as those trained with a noiseless channel. The results of this experiment are shown in Figure~\ref{fig:single_delay_ablation}. Note that in these experiments, we still enforce the same energy constraints on average across delays using the strategy proposed in Section~\ref{sec:delay_specialisation}.

The choice of delays that we train on has a significant impact on the performance of the network following training. Training on a single delay produces a high degree of specialisation in the case of delays of 0 (no delay), 1, 10, and 100, where the performance at that particular delay outperforms even the conditioned versions of the autoencoder, shown by the local peaks in the figure. This is a much less complex problem than the delay-conditioned scenarios and this local performance boost comes at the expense of a much poorer performance for all other delays.

Interestingly, we see that the performance in the case of training with a delay of 1000 produces a better average performance across a wider range of delays, indicating that the low-rate (worst case) strategy adopted is better able to generalise to the range of channel conditions (delays).

Decoder conditioning produces the smallest improvement in the performance. Encoder and encoder/decoder conditioning produce a much more marked improvement. This emphasises that the greatest performance gains are achieved through pre-coding against systematic distortions introduced by the delay using the encoder, but that the same effect can also be partially achieved using the decoder alone and a one-size-fits-all encoding strategy. The best average performance (slightly beating the encoder-only approach) is the encoder/decoder -conditioned approach. However, in practice, we expect that the delay may be unknown at the time of storage, but can be assumed to be known at the time of recovery.

\subsection{Energy Distribution}

\begin{figure*}
    \centering
    \subfloat[None (0.1J).]
    {
      \includegraphics[width=0.23\linewidth]{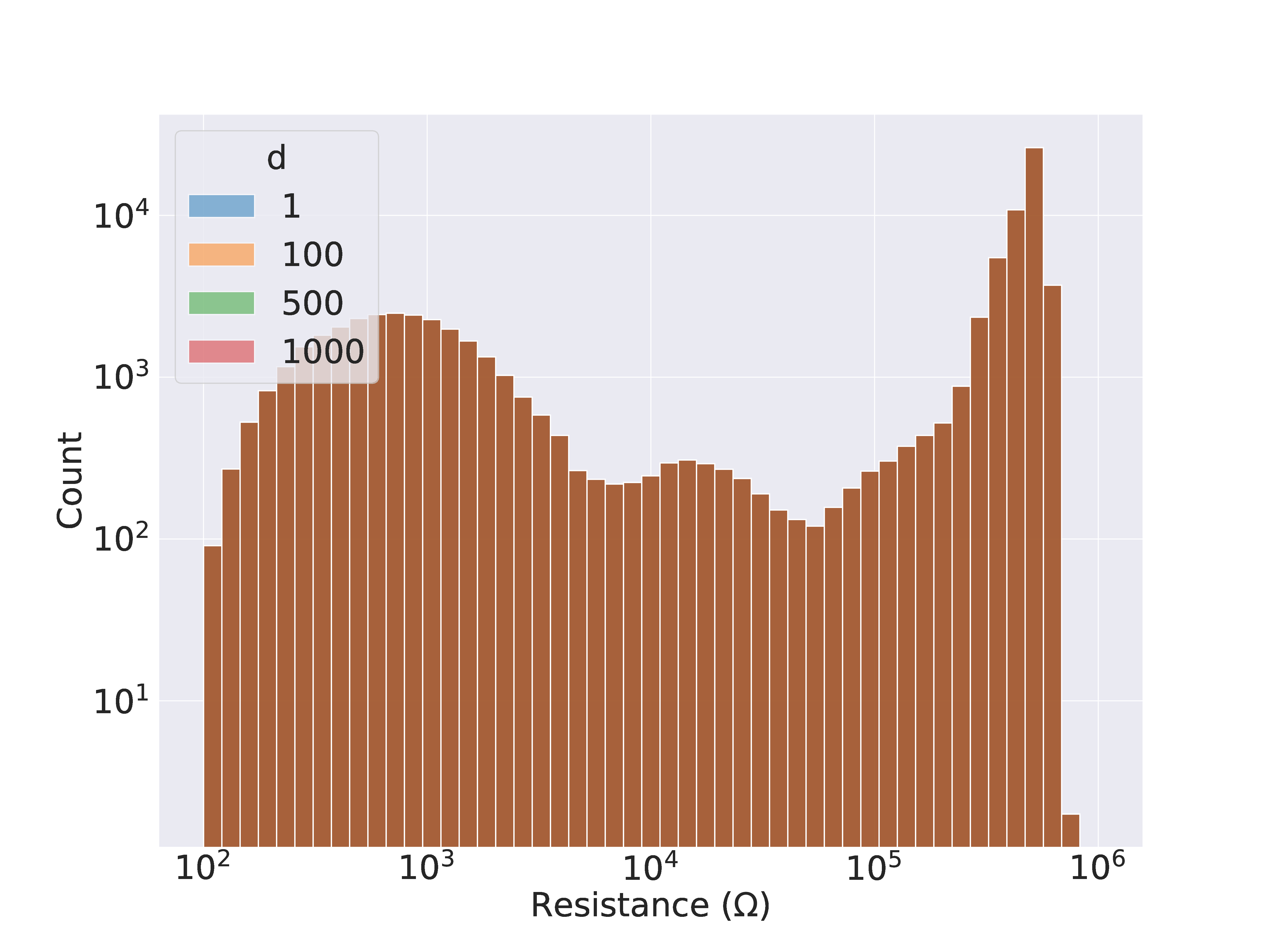}
      \label{fig:energy_0.1}
      }
      \hfill
    \subfloat[Encoder/decoder (0.1J).]{
      \includegraphics[width=0.23\linewidth]{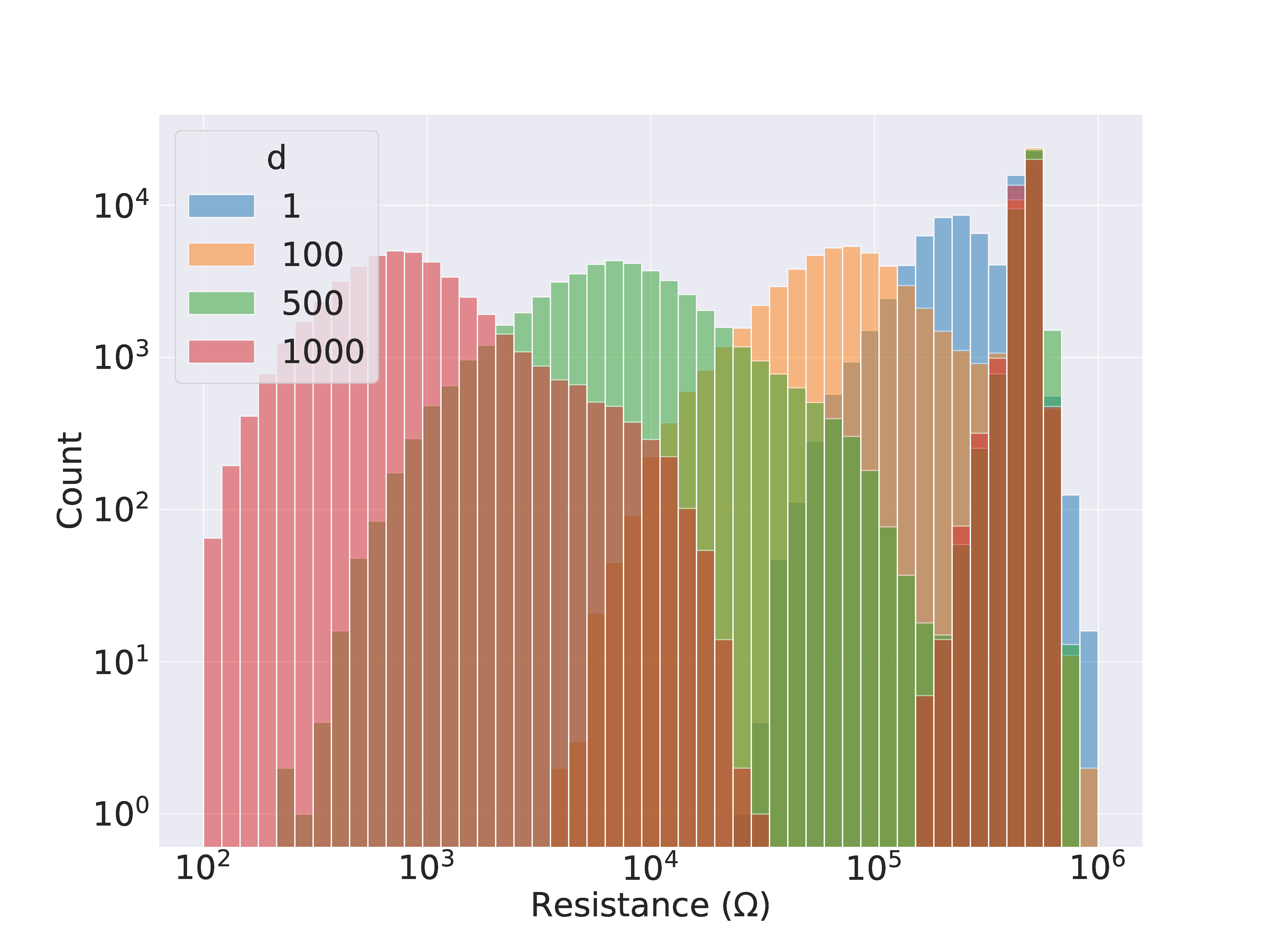}      \label{fig:energy_0.1_encoder_decoder}}
      \hfill
    \subfloat[Encoder (0.1J).]{
      \includegraphics[width=0.23\linewidth]{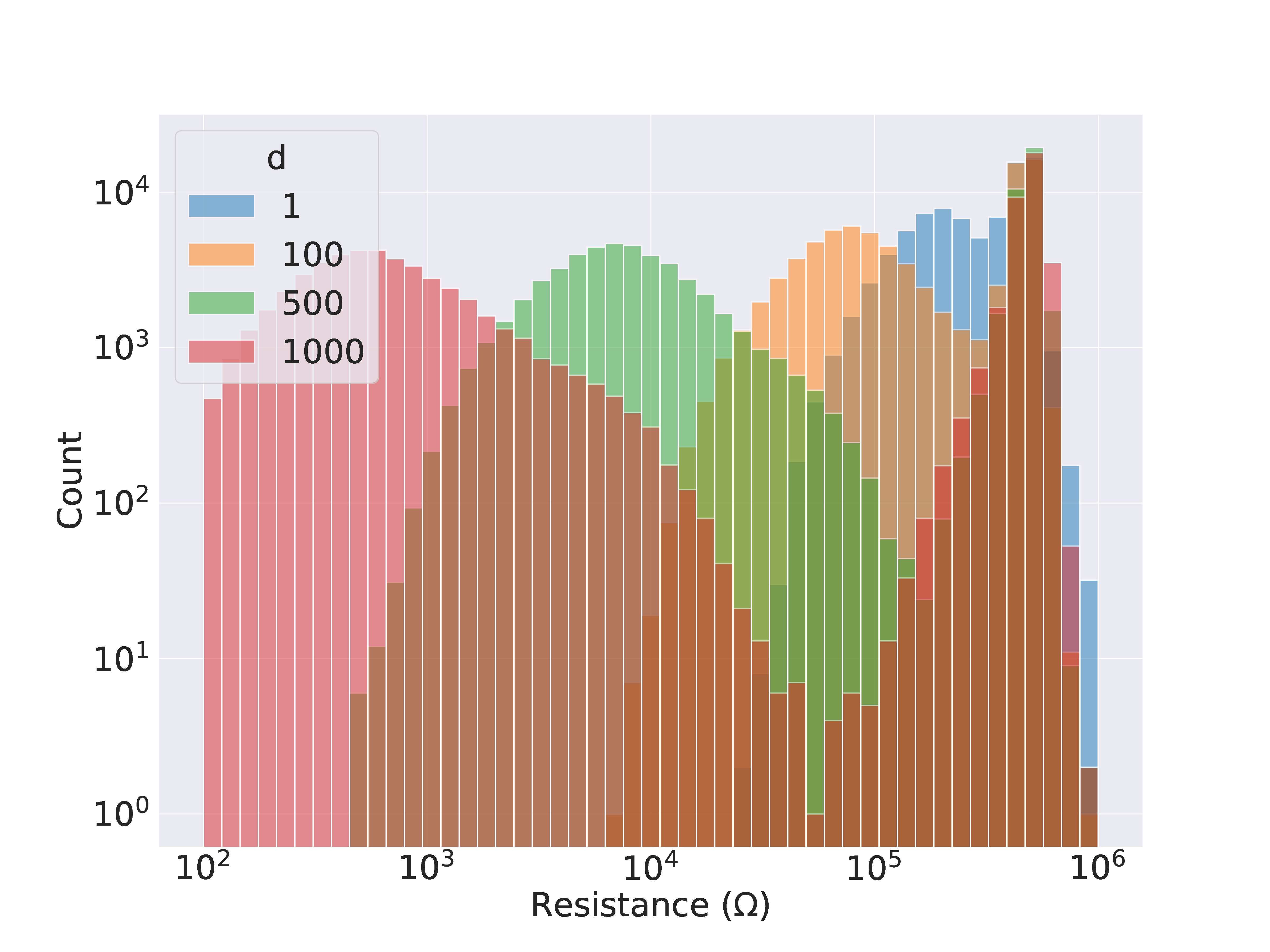}
    \label{fig:energy_0.1_encoder}}
    \hfill
    \subfloat[Decoder (0.1J).]{
      \includegraphics[width=0.23\linewidth]{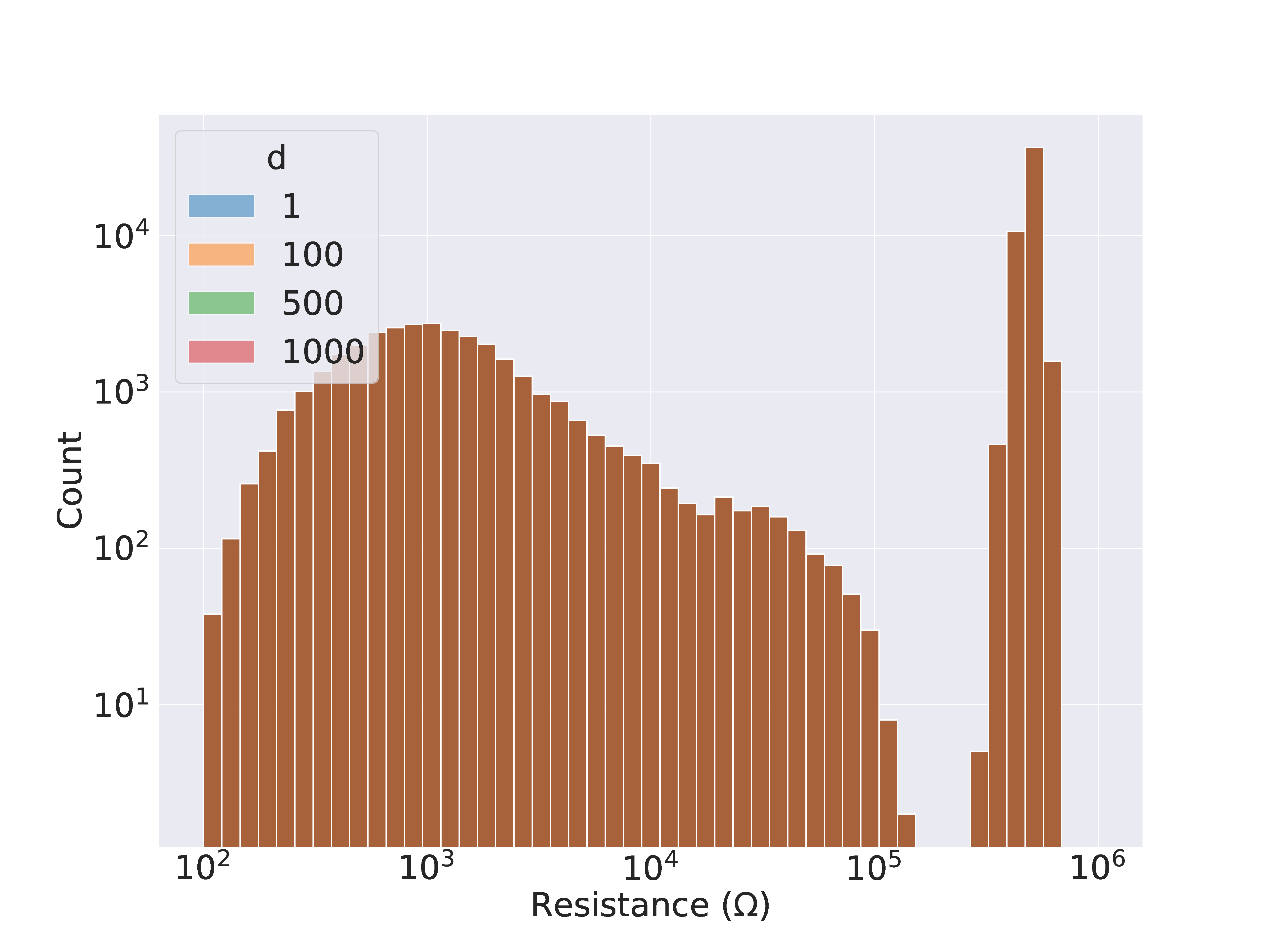}
      \label{fig:energy_0.1_decoder}
      }
    \hfill
    \subfloat[None (0.01J).]{
      \includegraphics[width=0.23\linewidth]{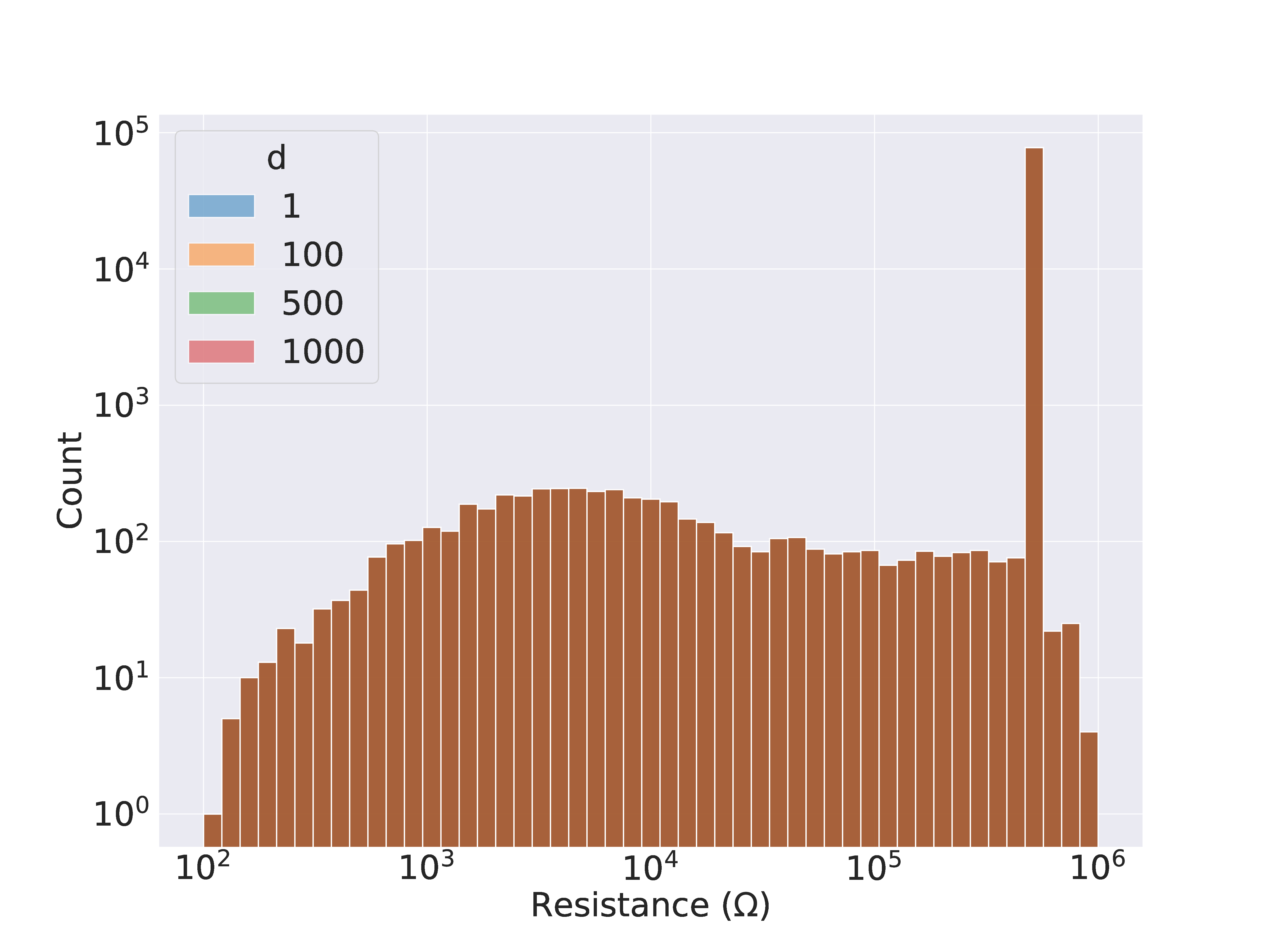}
      \label{fig:energy_0.01}}
    \hfill
    \subfloat[Encoder/decoder (0.01J).]{
      \includegraphics[width=0.23\linewidth]{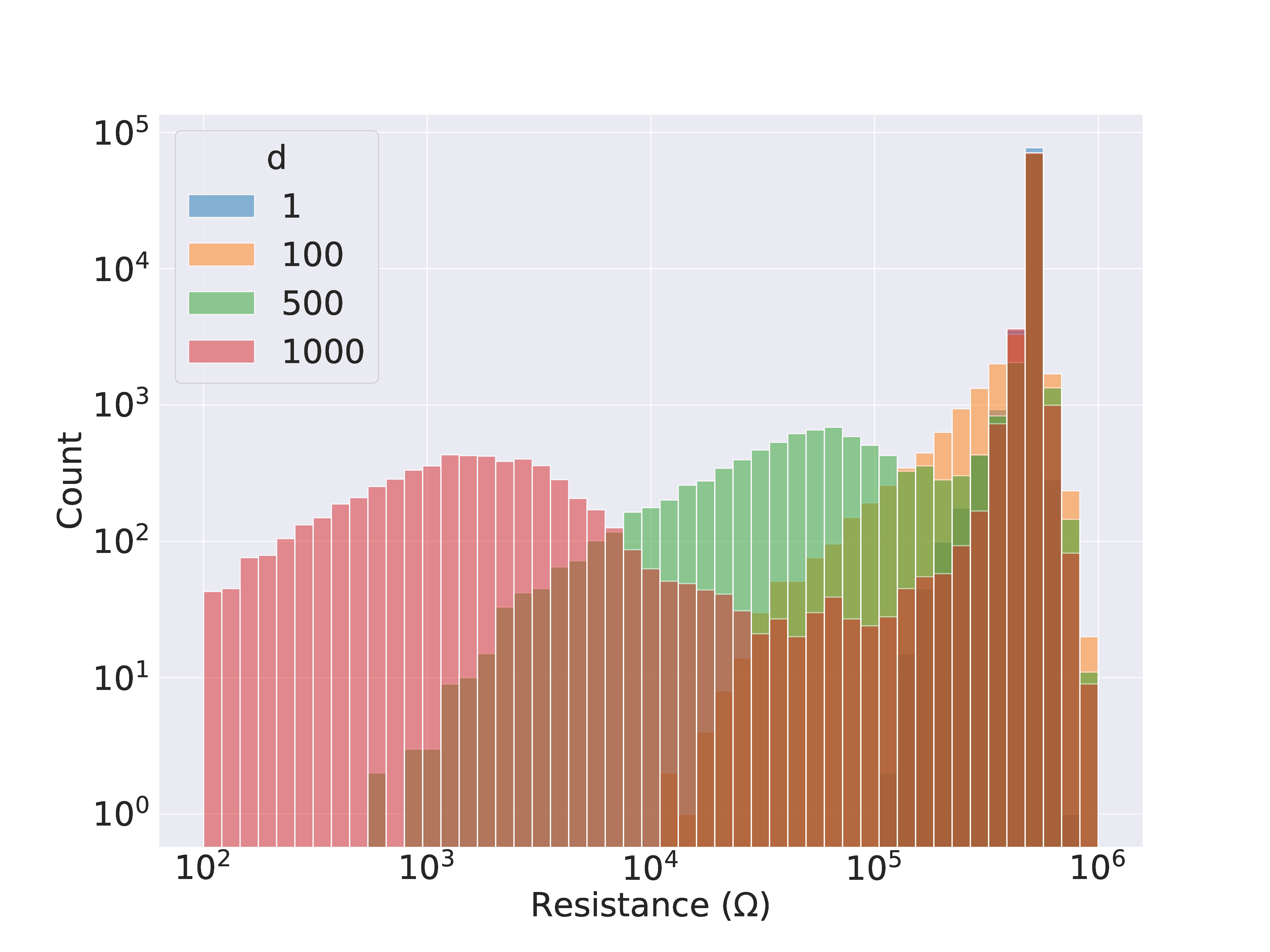}
      }
      \hfill
    \subfloat[Encoder (0.01J).]{
      \includegraphics[width=0.23\linewidth]{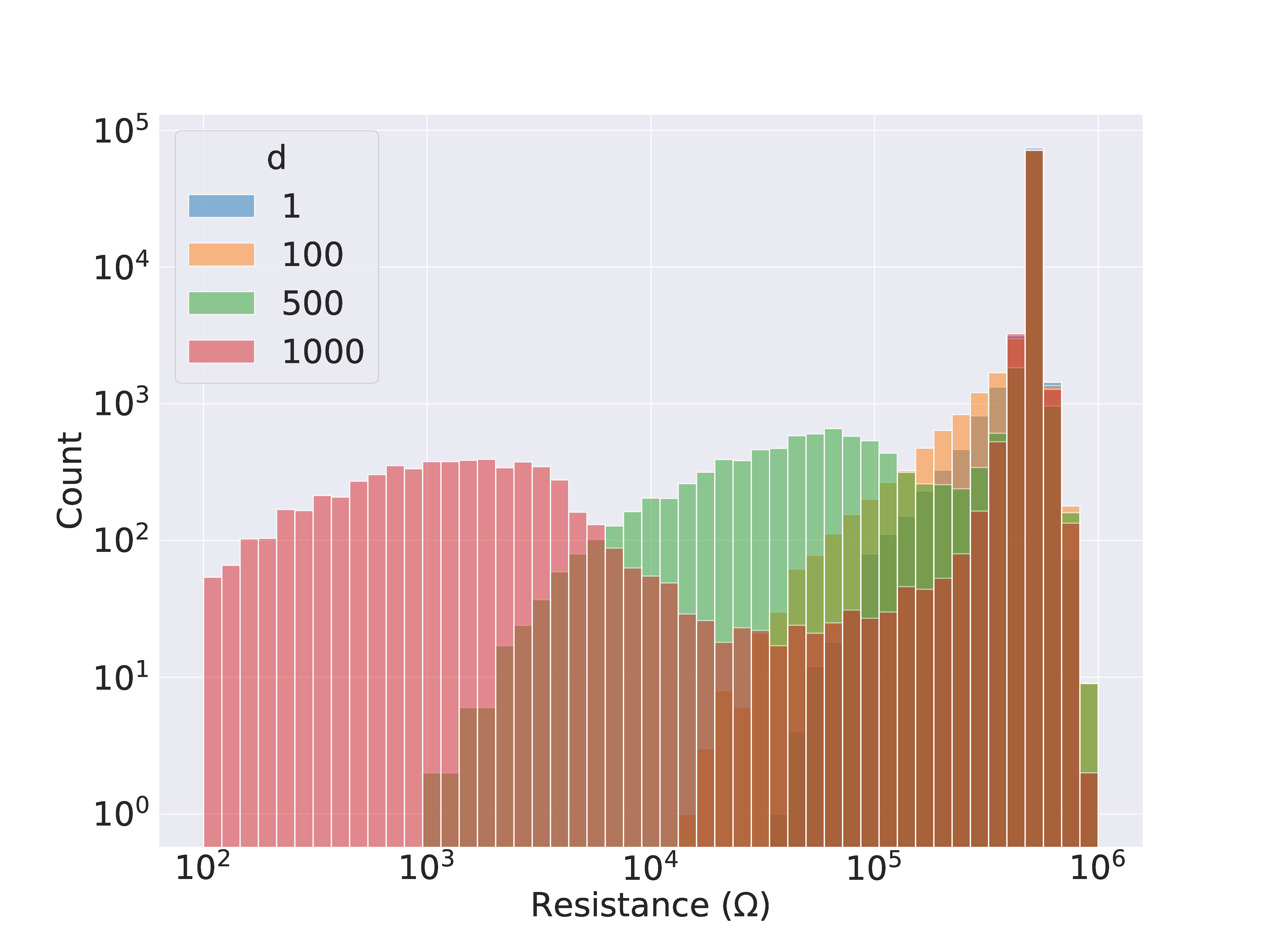}
      \label{fig:energy_0.01_encoder}}
      \hfill
    \subfloat[Decoder (0.01J).]{
      \includegraphics[width=0.23\linewidth]{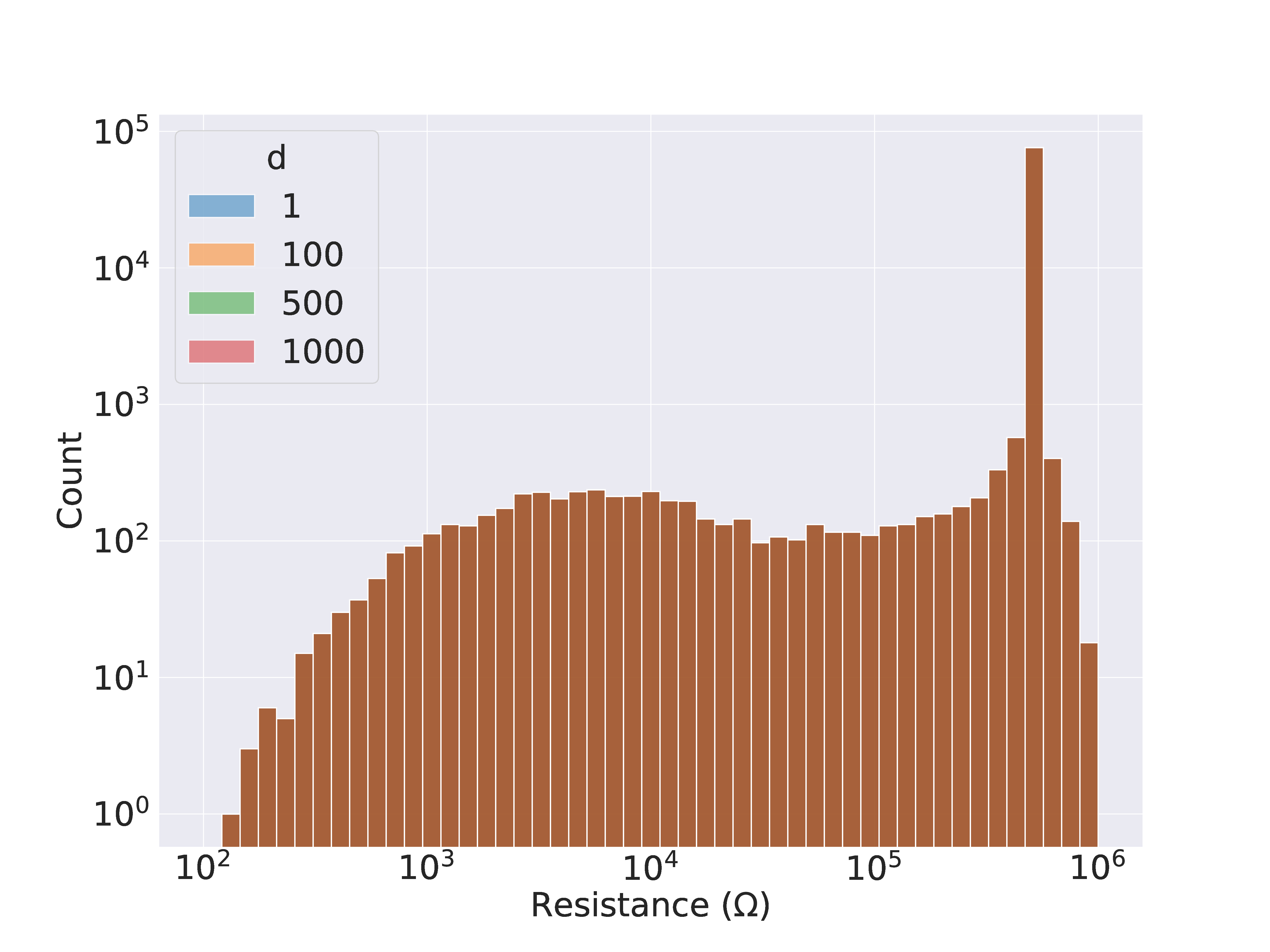}
      \label{fig:energy_0.01_decoder}}
    
    \caption{A comparison of the input distributions for the different encoder/decoder conditioned settings, for two different energy budgets: 0.1J, and 0.01J. We see that in the unconditioned cases, and the decoder conditioned cases, the histograms are the same across delays, whereas encoder and decoder conditioning leads to adaptation of the input distribution, with input distributions that are more costly in terms of the energy budget being allocated to larger, noisier, delay conditions. Note that the counts are shown on a logarithmic scale for improved clarity.}
    \label{fig:energy_hist_results}
\end{figure*}

Figure~\ref{fig:energy_hist_results} shows a comparison of the encoded distributions of (un-normalised) resistances for a variety of energy budgets and delay conditions in the encoder conditioned experiments. We see that using delay conditioning, the encoder is able to tailor the input distribution to the delay, in order to better distribute the available energy budget, and to improve performance for larger delays by increasing the power allocation, at the expense of reduced power allocation for lower delay settings, where there is less noise. Note that many histograms show peaks at the equilibrium resistance value. In such cases, we can hypothesise that the network chooses to use a sparse encoding strategy, where a few of the memristors are allocated a majority of the information content, along with a much higher energy budget.

Note that the delay conditioning allows the autoencoder to not only account for the different deterministic transformations introduced by different delays, but also to adopt different encoding schemes at different \ac{PSNR} levels. For example, it may choose to expend more energy on encoding information at a variety of lower, relatively stable resistance levels for small delays, under which conditions the states will be relatively stable. In contrast, in the high-delay setting, it may choose to instead expend energy on maximising the distance between a smaller number of discrete input symbols that will be more robust against more severe drift that occurs over a longer period of time.

\subsection{Ground Truth Model Evaluation}
\label{sec:real_dataset_evaluation}

In order to evaluate the efficacy of the \ac{cGAN} model for simulating the storage of information in the presence of resistive drift, we evaluate the autoencoder on the ground truth, event-based model, as a measure of its ability to generalise to the statistics of a real device, following training using a generative model. This also serves as a measure of the efficacy of the delay-conditioned \ac{cGAN} for simulating the true statistics of the ground truth model (in this case, the event-based model developed in \cite{el-geresyEventBasedSimulationStochastic2024}).

\begin{figure}
    \centering
    \includegraphics[width=\linewidth]{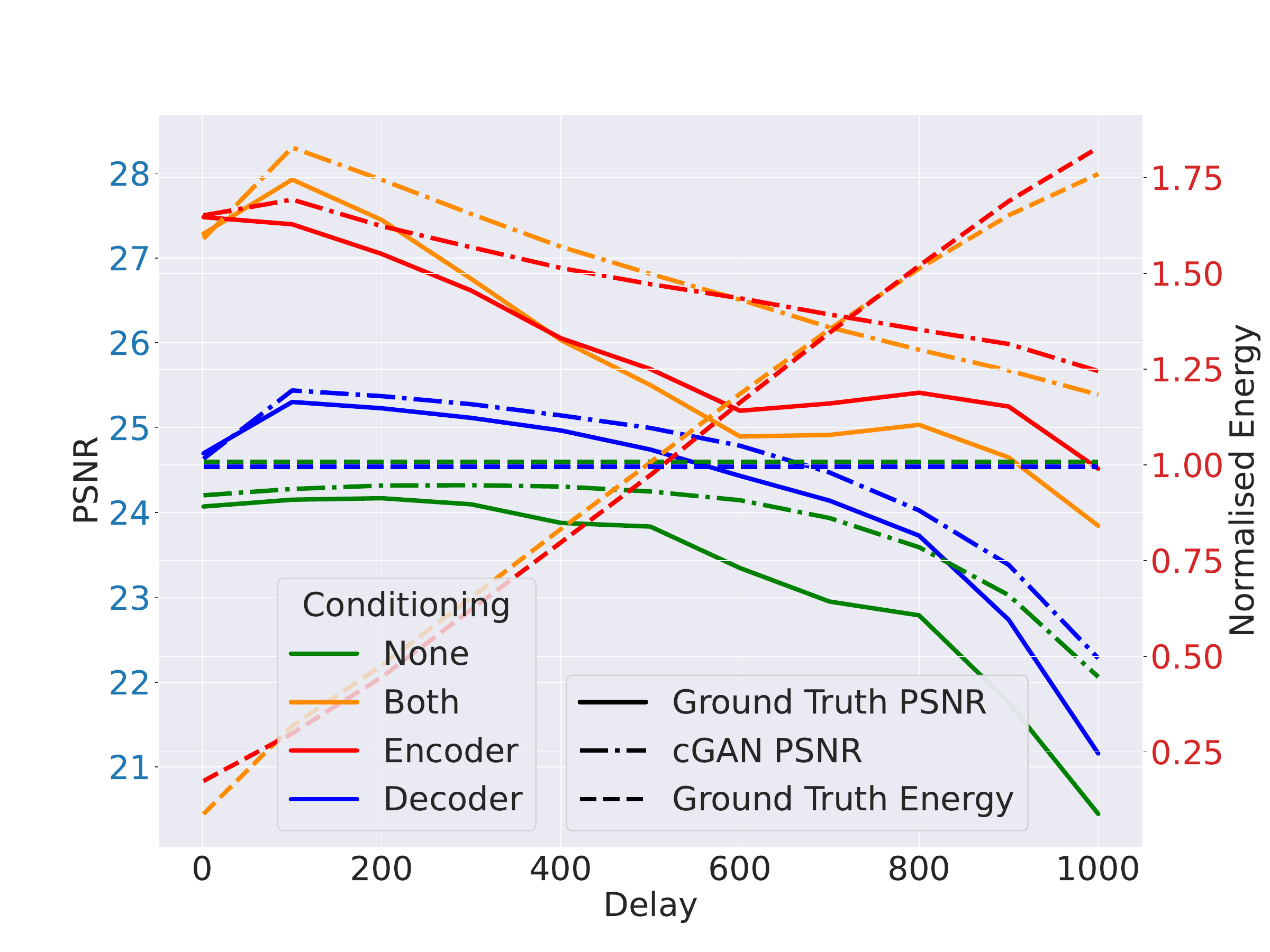}
    \caption{Evaluation on the ground truth event-based model for the reconstruction \ac{PSNR} of the autoencoders trained for an energy budget of 0.01J on the \ac{cGAN} model from \cite{el-geresyDelayConditionedGenerative2024}. We also show the energy consumption of the input distribution for each delay and input resistance condition. We plot the performance when evaluated on the \ac{cGAN} model alongside these results for comparison. }
    \label{fig:real_evaluation}
\end{figure}

Figure~\ref{fig:real_evaluation} shows the performance of the energy constrained autoencoders when evaluated on the ground truth event-based model, for the different delay-conditioning settings, for an energy budget of 0.01J and for delays in the range \([1, 1000]\), alongside an equivalent evaluation on the \ac{cGAN}. We see that for some delays, there is a drop in performance when we switch from evaluation on the trained \ac{cGAN} model to evaluation on the ground truth event-based model, highlighting regions where the statistics were not matched perfectly. However, the performance when evaluated on the ground truth model is generally comparable to that evaluated on the \ac{cGAN} model for most delays, demonstrating that the \ac{cGAN} is able to accurately model the conditional drift distribution, for a wide range of delay conditions, in order to enable the autoencoder to learn to store information in the presence of drift noise.

\section{Conclusion}
\label{sec:conclusion}

We have studied the problem of analogue information storage in memristive devices, and identified and characterised a novel trade-off between the energy expenditure and recovery quality.
Based on this trade-off, we have shown that a \ac{DeepJSCC} encoder/decoder pair can be trained on a differentiable model of the resistive drift in order to store images on an array of memristive devices, with a graceful trade-off between the number of devices used and the reconstruction error in terms of a chosen metric (such as the \ac{PSNR}). We demonstrated that conditioning on the delay could be achieved in order to enable the encoder and decoder to adapt the energy consumption and channel input distribution to different delay settings for improved performance, while still satisfying average power constraints.
To the best of our knowledge, this is the first time the concept of delay-dependent encoding/decoding has been introduced in the context of lossy data storage.

Future work could include modelling of, and \ac{JSCC} for, other forms of memristor noise including programming noise and noise arising from imperfect knowledge of the state due to accessing an array of memristors in a crossbar configuration. It could also include extension of the channel modelling and \ac{JSCC} techniques to stateful device models, through use of sequence generation architectures, such as transformers or recurrent architectures. Furthermore, the conditioning approaches developed could be investigated in terms of their ability to code against other forms of noise; in particular, the \ac{cGDN} architecture could be investigated as a tool for use in the context of other communication channels e.g. to condition on the channel \ac{SNR} in wireless channels.

\bibliography{references}
\bibliographystyle{ieeetr}

\end{document}